\documentclass[times,a4paper]{article}

\usepackage{fullpage}
\usepackage{amsmath}
\usepackage{amssymb}
\usepackage{graphicx}
\usepackage[numbers]{natbib}

\makeatletter
\usepackage{clrscode3e}
\usepackage{float}
\usepackage{rotating}
\usepackage{booktabs}
\usepackage{verbatim} % for comment environment

\graphicspath{{figs/}}

\newfloat{code}{tpbh}{aux}
\floatname{code}{Code}
\newcommand{\Foreach}{\kw{for each}\,\Indentmore}
\renewcommand{\For}{\kw{for}\,\Indentmore}

\title{On Disturbance State-Space Models and the Particle Marginal
  Metropolis-Hastings Sampler}

\author{Lawrence M. Murray\footnotemark[2]\
        \and Emlyn M. Jones \footnotemark[3]\
        \and John Parslow \footnotemark[3]}

\begin{document}

\maketitle

\renewcommand{\thefootnote}{\fnsymbol{footnote}}

\footnotetext[2]{CSIRO Mathematics, Informatics and Statistics, Perth,
  Australia ({\tt lawrence.murray@csiro.au}).}
\footnotetext[3]{CSIRO Marine and Atmospheric Research, Hobart, Australia.}

\begin{abstract}
We investigate nonlinear state-space models without a closed-form transition
density, and propose reformulating such models over their latent noise
variables rather than their latent state variables. In doing so the tractable
noise density emerges in place of the intractable transition density. For
importance sampling methods such as the auxiliary particle filter, this
enables importance weights to be computed where they could not be
otherwise. As case studies we take two multivariate marine biogeochemical
models and perform state and parameter estimation using the particle marginal
Metropolis-Hastings sampler. For the particle filter within this sampler, we
compare several proposal strategies over noise variables, all based on
lookaheads with the unscented Kalman filter. These strategies are compared
using conventional means for assessing Metropolis-Hastings efficiency, as well
as with a novel metric called the \emph{conditional acceptance rate} for
assessing the consequences of using an estimated, and not exact,
likelihood. Results indicate the utility of reformulating the model over noise
variables, particularly for fast-mixing process models.
\end{abstract}

%%%%
\section{Introduction}

\begin{figure}[tp]
\begin{minipage}[b]{0.48\textwidth}
\includegraphics[width=\textwidth]{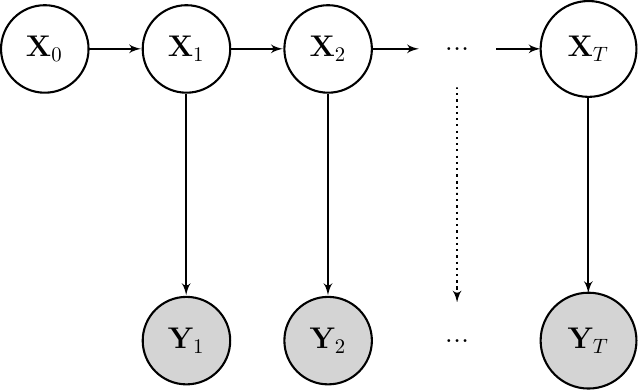}
\end{minipage}
\hfill
\begin{minipage}[b]{0.48\textwidth}
\includegraphics[width=\textwidth]{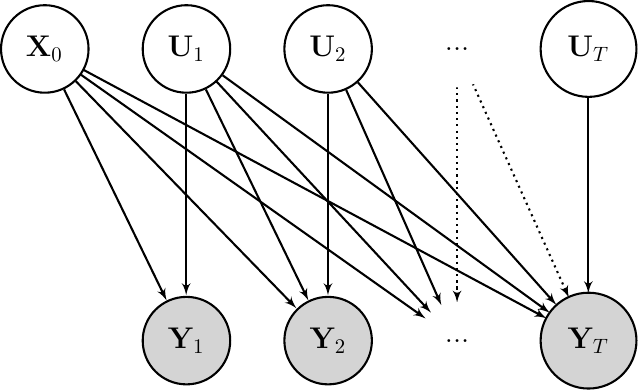}
\end{minipage}
\caption{Graphical models of \textbf{(a)} the conventional state-space model,
  and \textbf{(b)} the disturbance state-space model. The $\mathbf{U}_{1:T}$
  represent independent noise variables; conceptually, the disturbance
  state-space model simply reformulates the conventional model over these
  noise variables, the initial state, and the observations. Parameters,
  $\boldsymbol{\Theta}$, have been removed for clarity, but note that
  $\mathbf{U}_{1:T} \perp\!\!\!\perp \boldsymbol{\Theta}$, while all other
  variables depend on them.}
\label{fig:model}
\end{figure}

For $T$ time points, a sequence of observations
$\mathbf{y}_1,\ldots,\mathbf{y}_T$ of random variables
$\mathbf{Y}_1,\ldots,\mathbf{Y}_T \in \mathbb{R}^{N_y}$ is assumed given,
indicative of a latent initial condition $\mathbf{X}_0 \in \mathbb{R}^{N_x}$
and latent states $\mathbf{X}_1,\ldots,\mathbf{X}_T \in \mathbb{R}^{N_x}$. The
model is parameterised by static $\boldsymbol{\Theta} \in
\mathbb{R}^{N_\theta}$. The state transition is Markovian, and observations
are conditionally independent given the states. This is the conventional
\emph{state-space model}, which takes the form
\begin{equation}
p(\mathbf{y}_{1:T},\mathbf{x}_{0:T},\boldsymbol{\theta}) =
\left[\prod_{t=1}^T
  p(\mathbf{y}_t|\mathbf{x}_t,\boldsymbol{\theta})\right]
  \left[\prod_{t=1}^T p(\mathbf{x}_t|\mathbf{x}_{t-1},\boldsymbol{\theta})\right]
 p(\mathbf{x}_0|\boldsymbol{\theta}) p(\boldsymbol{\theta}),
\end{equation}
with the equivalent graphical model shown in Figure \ref{fig:model}(a).

The \emph{transition density},
$p(\mathbf{x}_t|\mathbf{x}_{t-1},\boldsymbol{\theta})$, may not have a closed
form, or if it does, it may be too expensive to compute. This has been noted
for diffusion processes~\citep{Beskos2006,Fearnhead2008}, and in fields such
as biochemistry~\citep{Golightly2008,Golightly2011}, Functional Magnetic
Resonance Imaging (fMRI)~\citep{Murray2011b}, and marine
biogeochemistry~\citep{Parslow2013}. In such cases it may be worth
reformulating the model over its latent noise variables rather than its latent
state variables. This can be done by explicitly introducing noise variables,
$\mathbf{U}_{1:T} \in \mathbb{R}^{N_u}$, in place of $\mathbf{X}_{1:T}$. These
may reflect, for example, independent Gaussian noise, or, in the most reduced
form, the emissions of a pseudorandom number generator. The model can then be
written:
\begin{equation}
p(\mathbf{y}_{1:T},\mathbf{u}_{1:T},\mathbf{x}_0,\boldsymbol{\theta}) =
\left[\prod_{t=1}^T
  p(\mathbf{y}_t|\mathbf{u}_{1:t},\mathbf{x}_0,\boldsymbol{\theta})\right]
  \left[\prod_{t=1}^T p(\mathbf{u}_t)\right]
 p(\mathbf{x}_0|\boldsymbol{\theta}) p(\boldsymbol{\theta}).
\end{equation}
The equivalent graphical model is shown in Figure \ref{fig:model}(b). We call
this the \emph{disturbance state-space model}. The conventional state-space
model is recovered from it by introducing a deterministic function
$f_\theta(\mathbf{u}_t,\mathbf{x}_{t-1}) \rightarrow \mathbf{x}_t$, permitting
the recursive recovery of a state trajectory $\mathbf{x}_{1:t}$ from any
sample $\{\mathbf{u}_{1:t},\mathbf{x}_0\}$.

In the disturbance state-space model the target becomes the posterior
distribution over the random variables
$\{\mathbf{U}_{1:T},\mathbf{X}_0,\boldsymbol{\Theta}\}$, rather than the
typical set $\{\mathbf{X}_{0:T},\boldsymbol{\Theta}\}$, and the tractable
\emph{noise density}, $p(\mathbf{u}_t)$, emerges in place of the intractable
transition density,
$p(\mathbf{x}_t|\mathbf{x}_{t-1},\boldsymbol{\theta})$. The noise density is
typically prescribed as having some simple parametric form, often Gaussian,
with the $\mathbf{U}_{1:T}$ independent of each other and of other
variables. This facilitates the design of proposal distributions when
sampling, such as within an auxiliary particle filter, in cases where this is
not possible under the conventional state-space model representation.

%In designing these proposal distributions, one limitation is
%that the number of noise variables, $N_u$, must be known \textsl{a priori},
%such that an appropriate number of variates can be generated. While \textsl{a
%  priori} knowledge of $N_u$ is common, it is not certain. Consider the
%numerical integration of a stochastic differential equation using an adaptive
%time step (e.g. \citep{Murray2011b}); the number of Wiener process increments
%to be simulated is only known at completion.

%Or consider a complex process model that is rejection sampled, with an
%unknown number of rejections to occur before acceptance [ref?].
%
% ^ if it is rejection sampled though, it's transition density would be closed
% form?

If $f_\theta(\cdot)$ is one-to-one, a change of variables for the transition
density might be considered:
\begin{eqnarray}
p(\mathbf{x}_t|\mathbf{x}_{t-1},\boldsymbol{\theta}) &=&
p\left(f_\theta^{-1}\left(\mathbf{x}_t\right)|\mathbf{x}_{t-1},\boldsymbol{\theta}\right) \left| \frac{df_\theta^{-1}(\mathbf{x}_t)}{d\mathbf{x}_t}
\right| \\
&=& p(\mathbf{u}_t) \left| \frac{df_\theta^{-1}(\mathbf{x}_t)}{d\mathbf{x}_t}
\right|.\label{eqn:change-of-variable}
\end{eqnarray}
When sampling from, say, $p(\mathbf{x}_t | \mathbf{x}_{t-1},
\boldsymbol{\theta}, \mathbf{y}_{1:t})$, one could likewise introduce an
importance proposal $q(\mathbf{x}_t|\mathbf{x}_{t-1},\boldsymbol{\theta}) =
q(\mathbf{u}_t) |df_\theta^{-1}(\mathbf{x}_t)/d\mathbf{x}_t|$. In computing
the importance weight, the Jacobian terms would cancel in the ratio, leaving
$p(\mathbf{u}_t)/q(\mathbf{u}_t)$. Because this must formally assume that
$f_\theta(\cdot)$ is one-to-one in order that the inverse
$f^{-1}_\theta(\cdot)$ exists, however, we prefer the more-general approach of
using the disturbance state-space model from the outset.

Reformulating to noise variables differs from the separation of tractable and
intractable components in the Rao-Blackwellisation of state-space models
\citep{Doucet2000b}, where it is typical to admit dependence of the tractable
(linear) component on the intractable (nonlinear) component, but not
vice-versa; in the disturbance state-space model, the intractable (nonlinear)
component depends on the tractable (noise) component.

\citet{Roberts2001} consider a similar idea to reparameterise a partially
observed diffusion process for the purposes of Gibbs sampling. The advantage
of doing so is that the conditional update of
$\boldsymbol{\Theta}|\mathbf{u}_{1:T},\mathbf{x}_0,\mathbf{y}_{1:T}$ is less
constrained than that of
$\boldsymbol{\Theta}|\mathbf{x}_{0:T},\mathbf{y}_{1:T}$, improving the mixing
of the sampler. In this work, however, a Metropolis-Hastings rather than Gibbs
update of $\boldsymbol{\Theta}$ is used, and no such advantage is conveyed. We
pick up on this point in the discussion (\S\ref{sec:discussion}).

There are alternative approaches to treat the absence of a closed-form
transition density. One option is linearisation, such as an
Euler-Maruyama~\cite{Golightly2008,Fearnhead2008,Golightly2011} or local
linearisation~\cite{Ozaki1993} of diffusion processes. This is not always
workable, however, owing to some processes being unstable when discretised
with low-order schemes. Higher-order or implicit discretisations are required
in such cases, yielding complicated closed-form expressions, if they can be
derived at all. A second option, supporting higher-order and implicit
discretisations, is to simply choose a proposal distribution that cancels
appearances of the transition density in weight
evaluations~\cite{Murray2011b}. The change-of-variables formulation above
might be seen as an extension of this, cancelling just the intractable
Jacobian term rather than the whole transition density. A third option is to
unbiasedly estimate the transition density, if possible, as in the
random-weight particle filter~\citep{Fearnhead2008}.

%The disturbance state-space model formulation has some limitations. Firstly,
%the number of random variates to be drawn in simulating the process model
%forward must be known \textsl{a priori}. While this is often the case, there
%are scenarios where it is not. Consider, for example, the numerical
%integration of a stochastic differential model using an adaptive time-step
%method~\cite{Murray2011b}, where the number of steps to traverse an interval
%of time, and thus the number of Wiener process increments to be drawn, is
%unknown upfront. Likewise, consider a process model which is sampled by
%rejection [cite example?], where the number of steps until acceptance is also
%unknown upfront. Secondly, we should not imagine that the design of proposal
%distributions over noise variables is necessarily a simple task. Given the
%mapping $f_\theta : \{\mathbf{U}_t,\mathbf{X}_{t-1}\} \rightarrow
%\mathbf{X}_t$ is likely to be highly nonlinear, the posterior distribution
%over $\mathbf{U}_t$ may be highly skewed.

The use of the disturbance state-space model in a particle filtering context is
given in \S\ref{sec:pf} along with a suite of proposal configurations based on
the unscented Kalman filter (UKF). We then turn to parameter estimation with
the particle marginal Metropolis-Hastings (PMMH) sampler in \S\ref{sec:pmmh},
and introduce the criterion of \emph{conditional acceptance rate} (CAR) to
compare the performance of the various configurations. Two case studies in
marine biogeochemistry are given in \S\ref{sec:cases} with extensive empirical
results. Concluding discussion appears in \S\ref{sec:discussion} and
\S\ref{sec:conclusion}.

%%%%
\section{The auxiliary particle filter for the disturbance state-space
  model}\label{sec:pf}

For given $\{\mathbf{x}_0,\boldsymbol{\theta}\}$, consider the joint filter
density of the conventional state-space model:
\begin{equation}
p(\mathbf{x}_{t-1:t}|\mathbf{x}_0,\boldsymbol{\theta},\mathbf{y}_{1:t}) \propto
  p(\mathbf{y}_t|\mathbf{x}_t,\boldsymbol{\theta})p(\mathbf{x}_t|\mathbf{x}_{t-1},\boldsymbol{\theta})p(\mathbf{x}_{t-1}|\mathbf{x}_0,\boldsymbol{\theta},\mathbf{y}_{1:t-1}).
\label{eqn:conventional-filter-density}
\end{equation}
The analogue in the disturbance state-space model is
\begin{equation}
p(\mathbf{u}_t,\mathbf{x}_{t-1}|\mathbf{x}_0,\boldsymbol{\theta},\mathbf{y}_{1:t})
\\ \propto p(\mathbf{y}_t|\mathbf{u}_t,\mathbf{x}_{t-1}, \boldsymbol{\theta})
p(\mathbf{u}_t)
p(\mathbf{x}_{t-1}|\mathbf{x}_0,\boldsymbol{\theta},\mathbf{y}_{1:t-1}).\label{eqn:filter-density}
\end{equation}
Note that the complete noise history, $\mathbf{u}_{1:t}$, does not need to
appear, as it is sufficient to recursively update and store $\mathbf{x}_t =
f_{\theta}(\mathbf{u}_{t},\mathbf{x}_{t-1})$ as the filter progresses, with
$\mathbf{x}_0$ establishing the base case of the recursion.

The auxiliary particle filter (APF)~\cite{Pitt1999} is readily
modified to sample from this. At time $t - 1$, the APF maintains a set of $M$
particles $\mathbf{x}^{1:M}_{t-1}$ with associated weights $w^{1:M}_{t-1}$,
normalised where required as $\tilde{w}^m_{t-1} = w^m_{t-1}/\sum_{i=1}^M
w^i_{t-1}$. To advance to time $t$, an auxiliary \emph{lookahead} propagation
and weighting procedure~\citep{Lin2009} is used to produce \emph{stage-one}
weights $\omega_t^{1:M}$, again normalised where required as
$\tilde{\omega}^m_t = \omega^m_t/\sum_{i=1}^M \omega^i_t$. For each particle
$m$, an ancestor index $a^m_t$ is drawn from the categorical distribution
$\mathcal{C}(\omega_t^{1:M})$, commonly called \emph{resampling} (see
e.g. \citet{Gordon1993} or \citet{Kitagawa1996}). For the disturbance
state-space model, the ancestor $\mathbf{x}^{a^m_t}_{t-1}$ is then extended by
sampling $\mathbf{u}^m_t \sim q^m_t(\mathbf{u}^m_t)$, where $q^m_t(\cdot)$ is
some importance proposal for the $m$th particle, and setting $\mathbf{x}^m_t =
f_{\theta}(\mathbf{u}^m_{t},\mathbf{x}^m_{t-1})$. The particle's weight is
then updated with
\begin{equation}\label{eqn:weight}
w^m_t = \frac{p(\mathbf{y}_t|\mathbf{x}^m_t, \boldsymbol{\theta}) 
p(\mathbf{u}^m_t)}{q^m_t(\mathbf{u}^m_t)}
\cdot \frac{\tilde{w}^{a^m_t}_{t - 1}}{\tilde{\omega}^{a^m_t}_t}.
\end{equation}
This is called the \emph{stage-two} weight. Code \ref{code:generic}
gives this generic APF algorithm for the disturbance state-space model.

\begin{code}[tp]
\begin{codebox}
\Procname{$\proc{APF}(\mathbf{x}_0, \boldsymbol{\theta})$}
\li Initialise with $\mathbf{x}^m_0 = \mathbf{x}_0$ and $\tilde{w}^m_0 = 1/M$
for $m = 1,\ldots,M$.
\li \For $t = 1,\ldots,T$
\li   \Foreach $m \in \{1,\ldots,M\}$
\li     Compute stage-one weight $\omega^m_t$
      \End
\zi
\li   \Foreach $m \in \{1,\ldots,M\}$
\li     $a^m_t \sim \mathcal{C}(\omega^{1:M}_t)$ \Comment resample
\li     $\mathbf{u}^m_t \sim q^m_t(\mathbf{u}^m_t)$ \Comment propose
\li     $\mathbf{x}^m_t \leftarrow
f_{\theta}(\mathbf{u}^m_t,\mathbf{x}^{a^m_t}_{t-1})$ \Comment propagate
\li     $w^m_t \leftarrow \frac{p(\mathbf{y}_t|\mathbf{x}^m_t,\boldsymbol{\theta})
p(\mathbf{u}^m_t)}{q^m_t(\mathbf{u}^m_t)} \cdot
\frac{\tilde{w}^{a^m_t}_{t-1}}{\tilde{\omega}^{a^m_t}_t} $ \Comment weight, stage two
      \End
    \End
\end{codebox}

\caption{Generic auxiliary particle filter over a disturbance state-space
  model.}
\label{code:generic}
\end{code}

For the conventional state-space model, the locally optimal, or \emph{fully
  adapted} proposal, is $q^m_t(\mathbf{x}^m_t) \equiv
p(\mathbf{x}^m_t|\mathbf{x}^{a^m_t}_{t-1},\boldsymbol{\theta},\mathbf{y}_t)$~\cite{Pitt1999}. For
the disturbance state-space model, its analogue is $q^m_t(\mathbf{u}^m_t)
\equiv
p(\mathbf{u}^m_t|\mathbf{x}^{a^m_t}_{t-1},\boldsymbol{\theta},\mathbf{y}_t)$. These
cannot be derived analytically for the models of interest in this work,
however, and so attention is given to reasonable approximations instead. Three
such approximations are considered. The first scheme is the ordinary bootstrap
particle filter \citep{Gordon1993}. The second and third use the UKF in
different ways, the third similar to the existing unscented particle filter
\citep{VanDerMerwe2000} but adapted to the disturbance state-space model. Each
proposal scheme is tested both with and without a lookahead component for
computing non-trivial stage-one weights, giving six methods in total. All six
methods are detailed in this section and summarised in Table
\ref{tab:algorithms}.

\begin{table}
\footnotesize
\centering
\begin{tabular}{lllr}
\toprule
Method & Stage-one weight, $\omega^m_t$ & Proposal, $q^m_t$ & Number of propagations \\
\midrule
PF0 & $w^m_{t-1}$ & $p(\mathbf{u}_t)$ & $M$ \\
PF1 & $p(\mathbf{y}_t|\mathbf{x}_t = f_\theta(\mathbf{0}, \mathbf{x}^m_{t-1}),
\boldsymbol{\theta}) w^m_{t-1}$ & $p(\mathbf{u}_t)$ & $2M$ \\
MUPF0 & $w^m_{t-1}$ & $\hat{p}(\mathbf{u}_t|\mathbf{x}_0,\boldsymbol{\theta},\mathbf{y}_{1:t})$ & $M + 2(N_u + N_x + N_y) + 1$ \\
MUPF1 & $p(\mathbf{y}_t|\mathbf{x}_t = f_\theta(\hat{\boldsymbol{\mu}}_t,
\mathbf{x}^m_{t-1}), \boldsymbol{\theta}) w^m_{t-1}$ &
$\hat{p}(\mathbf{u}_t|\mathbf{x}_0,\boldsymbol{\theta},\mathbf{y}_{1:t})$ &
$2M + 2(N_u + N_x + N_y) + 1$ \\
CUPF0 & $w^m_{t-1}$ & $\hat{p}^m(\mathbf{u}^m_t|
\mathbf{x}^{a^m_t}_{t-1},\boldsymbol{\theta},\mathbf{y}_{1:t})$ & $2M(N_u +
N_y + 1)$ \\
CUPF1 & $\hat{p}^m(\mathbf{y}_t|\mathbf{x}^m_{t-1}, \boldsymbol{\theta})
w^m_{t-1}$ & $\hat{p}^m(\mathbf{u}^m_t|
\mathbf{x}^{a^m_t}_{t-1},\boldsymbol{\theta},\mathbf{y}_{1:t})$ & $2M(N_u +
N_y + 1)$\\
\bottomrule
\end{tabular}

\caption{Summary of the auxiliary particle filter proposal mechanisms
  explored. See \S\ref{sec:pf} for details.}
\label{tab:algorithms}
\end{table}

\subsection{Bootstrap particle filter}

The simplest approach, that of the bootstrap filter \citep{Gordon1993}, sets
\begin{equation} \label{eqn:pf-proposal}
q^m_t(\mathbf{u}^m_t) \equiv q_t(\mathbf{u}_t) \equiv p(\mathbf{u}_t),
\end{equation}
and
\begin{equation} \label{eqn:pf0-weight}
\omega^m_t = w^m_{t-1}.
\end{equation}
This is straightforward to apply, and indeed does not benefit from
reformulating the model over noise variables. Both the proposal and weights
cancel in (\ref{eqn:weight}), so that particles are simply weighted by the
likelihood $p(\mathbf{y}_t|\mathbf{x}^m_t,\boldsymbol{\theta})$. We denote
this method PF0.

Lacking $p(\mathbf{y}_t|\mathbf{x}_{t-1},\boldsymbol{\theta})$, an analytical
lookahead is not forthcoming. A deterministic single-point pilot lookahead,
simulating with $\mathbf{u}_t = \mathbf{0}$, can offer improvement in some
cases \citep{Lin2009}. By modifying the stage-one weights to
\begin{equation}
\omega^m_t = p(\mathbf{y}_t|\mathbf{x}_t = f_\theta(\mathbf{0}, \mathbf{x}^m_{t-1}),
\boldsymbol{\theta}) w^m_{t-1},\label{eqn:pf1-weight}
\end{equation}
we obtain a similar method with a lookahead, and denote it PF1.

\subsection{Marginal unscented particle filter}\label{sec:mupf}

We next attempt to draw on analytical approximations to inform the proposal
distribution. The particular focus is on the UKF, which, for modest state
sizes, tends to outperform~\citep{Wan2000} other approximate nonlinear Kalman
filtering approaches, such as the extended~\citep{Smith1962} and ensemble
\citep{Evensen1994,Evensen2009} variants. The UKF approximates the time
marginals $p(\mathbf{u}_t|\mathbf{x}_0,\boldsymbol{\theta},\mathbf{y}_{1:t})$
using a Gaussian distribution. We denote the approximation
$\hat{p}(\mathbf{u}_t|\mathbf{x}_0,\boldsymbol{\theta}, \mathbf{y}_{1:t})
\equiv \mathcal{N}(\hat{\boldsymbol{\mu}}_t, \hat{\Sigma}_t)$. At each time,
$2(N_u + N_x + N_y) + 1$ number of $\sigma$-points are crafted about the mean
of the Gaussian distribution, propagated through the process model and
specifically weighted to compute a Gaussian approximation to $p(\mathbf{u}_t,
\mathbf{y}_t|\mathbf{x}_0, \boldsymbol{\theta},
\mathbf{y}_{1:t-1})$. Conditioning this on the actual observed value
$\mathbf{y}_t$ delivers the approximate time marginal
$\hat{p}(\mathbf{u}_t|\mathbf{x}_0, \boldsymbol{\theta},
\mathbf{y}_{1:t})$. See \citet{Julier1997} and \citet{Wan2000} for details.

The first UKF-based approach adopted is to use the \textsl{marginal} UKF
approximations $\hat{p}(\mathbf{u}_t|\mathbf{x}_0, \boldsymbol{\theta},
\mathbf{y}_{1:t})$ at each time as a common proposal for each particle:
\begin{equation}\label{eqn:mupf-proposal}
q^m_t(\mathbf{u}^m_t) \equiv q_t(\mathbf{u}_t) \equiv
\hat{p}(\mathbf{u}_t|\mathbf{x}_0,\boldsymbol{\theta},\mathbf{y}_{1:t}) \equiv \mathcal{N}(\hat{\boldsymbol{\mu}}_t, \hat{\Sigma}_t).
\end{equation}

We call this the \emph{marginal unscented particle filter} (MUPF), described
in Code \ref{code:mupf}. By combining with the stage-one weights
(\ref{eqn:pf0-weight}) we have the MUPF0 method. With stage-one weights
\begin{equation}\label{eqn:mupf1-weight}
\omega^m_t = p(\mathbf{y}_t|\mathbf{x}_t = f_\theta(\hat{\boldsymbol{\mu}}_t,
\mathbf{x}^m_{t-1}), \boldsymbol{\theta}) w^m_{t-1}
\end{equation}
we have the MUPF1 method. Intuitively, using
$f_\theta(\hat{\boldsymbol{\mu}}_t, \mathbf{x}^m_{t-1})$ here seems more
appealing than the $f_\theta(\mathbf{0}, \mathbf{x}^m_{t-1})$ that appears in
(\ref{eqn:pf1-weight}).

Note that the same time marginal $\hat{p}(\mathbf{u}_{t} | \mathbf{x}_0,
\boldsymbol{\theta}, \mathbf{y}_{1:t})$ is used for each particle, not the
conditional for the $m$th particle, $\hat{p}^m(\mathbf{u}^m_{t} |
\mathbf{x}^{a^m_t}_{t-1}, \boldsymbol{\theta}, \mathbf{y}_{1:t})$, which is
the basis for the next scheme. While the conditional proposal is no doubt
preferable conceptually, the marginal proposal may be justifiable for
fast-mixing models, and additionally enables the computational advantage of
running the UKF offline from the particle filter. The overhead of the method
is slight, with only $2(N_u + N_x + N_y) + 1$ additional propagations for the
whole filter. This linear scaling with the number of dimensions is likely
small compared to $M$, the number of particles, which would typically scale
exponentially with the same.

%In the case where $\mathbf{U}_{1:T} \sim \text{i.i.d.}\,\mathcal{N}(0,1)$,
%the also-Gaussian marginals of the UKF afford some efficiencies in the weight
%calculation after slotting (\ref{eqn:mupf0-proposal}) or
%(\ref{eqn:mupf1-proposal}) into (\ref{eqn:weight}). These are given in
%\S\ref{sec:gaussian-weight}. This also suggests that, even in the presence of
%a closed-form transition density over state variables, a reformulation of the
%state-space model to noise variables may still be of benefit,
%computationally, to speed up weight calculations.

\begin{code}[tp]
\begin{codebox}
\Procname{$\proc{MUPF}(\mathbf{x}_0, \boldsymbol{\theta})$}
\li Initialise with $\mathbf{x}^m_0 = \mathbf{x}_0$ and $\tilde{w}^m_0 = 1/M$
for $m = 1,\ldots,M$.
\li Run a UKF to produce filtering densities
$\hat{p}(\mathbf{u}_t|\mathbf{x}_0,\boldsymbol{\theta},\mathbf{y}_{1:t})
\equiv \mathcal{N}(\hat{\boldsymbol{\mu}}_t,\hat{\Sigma}_t)$,
for $t = 1,\ldots,T$.
\zi
\li \For $t = 1,\ldots,T$
\li   \Foreach $m \in \{1,\ldots,M\}$
\li     \If doing MUPF1 \Then
\li       $\hat{\mathbf{x}}^m_t \leftarrow
f_{\theta}(\hat{\boldsymbol{\mu}}_t, \mathbf{x}^m_{t-1})$ \Comment
look-ahead
\li       $\omega^m_t \leftarrow
p(\mathbf{y}_t|\hat{\mathbf{x}}^m_t,\boldsymbol{\theta}) w^m_{t-1}$ \Comment
weight, stage one
\li     \Else doing MUPF0
\li       $\omega^m_t \leftarrow w^m_{t-1}$ \Comment weight, stage one
        \End
      \End
\zi
\li   \Foreach $m \in \{1,\ldots,M\}$
\li     $a^m_t \sim \mathcal{C}(\omega^{1:M}_t)$ \Comment resample
\li     $\mathbf{u}^m_t \sim \mathcal{N}(\hat{\boldsymbol{\mu}}_t,\hat{\Sigma}_t)$ \Comment propose
\li     $\mathbf{x}^m_t \leftarrow
f_{\theta}(\mathbf{u}^m_t,\mathbf{x}^{a^m_t}_{t-1})$ \Comment propagate
\li     $w^m_t \leftarrow \frac{p(\mathbf{y}_t|\mathbf{x}^m_t,\boldsymbol{\theta})
p(\mathbf{u}^m_t)}{q^m_t(\mathbf{u}^m_t)} \cdot
\frac{\tilde{w}^{a^m_t}_{t-1}}{\tilde{\omega}^{a^m_t}_t} $ \Comment weight, stage two
      \End
    \End
\end{codebox}

\caption{Marginal unscented particle filter (MUPF).}
\label{code:mupf}
\end{code}

\subsection{Conditional unscented particle filter}\label{sec:cupf}

Finally, by conditioning the lookahead for each particle on the state of that
particle, we arrive at the \emph{conditional unscented particle filter}
(CUPF), detailed in Code \ref{code:cupf}. It is similar to the unscented
particle filter~\citep{VanDerMerwe2000} but modified for the disturbance
state-space model. The proposal is:
\begin{equation}\label{eqn:cupf-proposal}
q^m_t(\mathbf{u}^m_t) \equiv \hat{p}^m(\mathbf{u}^m_t|\mathbf{x}^{a^m_t}_{t-1},
\boldsymbol{\theta}, \mathbf{y}_{1:t}) \equiv
\mathcal{N}(\hat{\boldsymbol{\mu}}^m_t, \hat{\Sigma}^m_t).
\end{equation}
A UKF is run for each particle at each time step to construct this proposal
distribution. Each UKF requires $2(N_u + N_y) + 1$ number of $\sigma$-point
propagations\footnote{Conditioning on $\mathbf{x}_{t-1}$ removes $N_x$ from
  the dimensionality of the unscented transformation, so it does not appear
  here.}, in addition to the subsequent propagation of the particle
itself. This means $2M(N_u + N_y + 1)$ propagations for the whole filter.
Combined with the stage-one weights (\ref{eqn:pf0-weight}) we have the CUPF0
method. The alternative: rather than a single-point pilot lookahead,
substantial improvement might be had by using the likelihood, marginalised
over $\mathbf{u}_t$, that is approximated by the UKF:
\begin{equation}\label{eqn:cupf1-weight}
\omega^m_t = \hat{p}^m(\mathbf{y}_t|\mathbf{x}^m_{t-1}, \boldsymbol{\theta})
w^m_{t-1}.
\end{equation}
Use of these weights gives the CUPF1 method, which requires very little
additional computation over CUPF0.

\begin{code}[tp]
\begin{codebox}
\Procname{$\proc{CUPF}(\mathbf{x}_0, \boldsymbol{\theta})$}
\li Initialise with $\mathbf{x}^m_0 = \mathbf{x}_0$ and $\tilde{w}^m_0 = 1/M$
\li \For $t = 1,\ldots,T$
\li   \Foreach $m \in \{1,\ldots,M\}$
\li       Run a UKF to produce
$\hat{p}^m(\mathbf{u}_t|\mathbf{x}^m_{t-1},\boldsymbol{\theta},\mathbf{y}_{1:t})
\equiv \mathcal{N}(\hat{\boldsymbol{\mu}}^m_t,\hat{\Sigma}^m_t)$.
\li     \If doing CUPF1 \Then
\li       $\omega^m_t \leftarrow
\hat{p}^m(\mathbf{y}_t|\mathbf{x}^m_{t-1},\boldsymbol{\theta}) w^m_{t-1}$ \Comment
weight, stage one
\li     \Else doing CUPF0
\li       $\omega^m_t \leftarrow w^m_{t-1}$ \Comment weight, stage one
        \End
      \End
\zi
\li   \Foreach $m \in \{1,\ldots,M\}$
\li     $a^m_t \sim \mathcal{C}(\omega^{1:M}_t)$ \Comment resample
\li     $\mathbf{u}^m_t \sim \mathcal{N}(\hat{\boldsymbol{\mu}}^m_t,\hat{\Sigma}^m_t)$ \Comment propose
\li     $\mathbf{x}^m_t \leftarrow
f_{\theta}(\mathbf{u}^m_t,\mathbf{x}^{a^m_t}_{t-1})$ \Comment propagate
\li     $w^m_t \leftarrow \frac{p(\mathbf{y}_t|\mathbf{x}^m_t,\boldsymbol{\theta})
p(\mathbf{u}^m_t)}{q^m_t(\mathbf{u}^m_t)} \cdot
\frac{\tilde{w}^{a^m_t}_{t-1}}{\tilde{\omega}^{a^m_t}_t} $ \Comment weight, stage two
      \End
    \End
\end{codebox}

\caption{Conditional unscented particle filter (CUPF).}
\label{code:cupf}
\end{code}

%%%%
\section{The particle marginal Metropolis-Hastings sampler for the disturbance
  state-space model}\label{sec:pmmh}

For joint state and parameter estimation we target the posterior density
$p(\mathbf{u}_{1:T},\mathbf{x}_0,\boldsymbol{\theta}|\mathbf{y}_{1:T})$. This
can be factorised as either:
\begin{equation}\label{eqn:pi1}
p_1(\mathbf{u}_{1:T},\mathbf{x}_0|\boldsymbol{\theta},\mathbf{y}_{1:T})p_2(\boldsymbol{\theta}|\mathbf{y}_{1:T})
\end{equation}
or
\begin{equation}\label{eqn:pi2}
p_1(\mathbf{u}_{1:T}|\mathbf{x}_0,\boldsymbol{\theta},\mathbf{y}_{1:T})p_2(\mathbf{x}_0,\boldsymbol{\theta}|\mathbf{y}_{1:T}).
\end{equation}
In either case, the first factor, $p_1(\cdot)$, is targeted using an APF,
described in \S\ref{sec:pf}. The second factor, $p_2(\cdot)$, is targeted in
an outer loop around the particle filter using Metropolis-Hastings (MH)
\citep{Metropolis1953,Hastings1970}. The particle filter nested within MH
defines the particle marginal Metropolis-Hastings (PMMH) sampler, from the
family of particle Markov chain Monte Carlo (PMCMC)
methods~\citep{Andrieu2010}.

Factorisation (\ref{eqn:pi1}) requires that a good importance proposal is
available for the sampling of $\mathbf{X}_0$ in the particle filter. If a good
proposal is not available, the sample weights will be degenerate. In such
cases (\ref{eqn:pi2}) is the more attractive set up. It replaces the
importance sample of $\mathbf{X}_0$ with local MH moves, which are typically
easier to design. The factorisation (\ref{eqn:pi2}) is used below.

In the outer loop, a proposed move from $\{\mathbf{x}_0,\boldsymbol{\theta}\}$
to $\{\mathbf{x}'_0,\boldsymbol{\theta}'\} \sim
\rho(\mathbf{x}'_0,\boldsymbol{\theta}'|\mathbf{x}_0,\boldsymbol{\theta})$ is
accepted with probability
\begin{equation}
\min\left[1, \frac{p(\mathbf{y}_{1:T}|\mathbf{x}'_0,\boldsymbol{\theta}')p(\mathbf{x}'_0,\boldsymbol{\theta}')\rho(\mathbf{x}_0,\boldsymbol{\theta}|\mathbf{x}'_0,\boldsymbol{\theta}')}{p(\mathbf{y}_{1:T}|\mathbf{x}_0,\boldsymbol{\theta})p(\mathbf{x}_0,\boldsymbol{\theta})\rho(\mathbf{x}'_0,\boldsymbol{\theta}'|\mathbf{x}_0,\boldsymbol{\theta})}\right],
\end{equation}
where $\rho(\cdot)$ is a proposal distribution over parameters, and the
marginal likelihoods $p(\mathbf{y}_{1:T}|\mathbf{x}'_0,\boldsymbol{\theta}')$
are estimated by an APF targeting $p_1(\cdot)$ in (\ref{eqn:pi2}). The
estimator is~\cite{DelMoral2004,Andrieu2010}:
\begin{equation}\label{eqn:likelihood}
p(\mathbf{y}_{1:T}|\mathbf{x}_0,\boldsymbol{\theta}) \approx \prod_{t=1}^T
\left[ \frac{1}{M}\sum_{m=1}^M w^m_t\right]
\end{equation}
This assumes that normalised stage-one weights are used in computing stage-two
weights, as in the preceding introduction. A proof of unbiasedness is given in
\citet{DelMoral2004}.
%in which case one replacess (\ref{eqn:weight}) with
%\begin{equation}
%\tilde{w}^m_t = \frac{p(\mathbf{y}_t|\mathbf{x}^m_t, \boldsymbol{\theta}) 
%p(\mathbf{u}^m_t)}{q^m_t(\mathbf{u}^m_t)}
%\cdot \frac{w^{a^m_t}_{t - 1}}{\omega^{a^m_t}_t/\sum_{m=1}^M\omega^m_t},
%\end{equation}
%and the likelihood estimator is
%\begin{eqnarray}
%p(\mathbf{y}_{1:T}|\mathbf{x}_0,\boldsymbol{\theta}) &\approx& \prod_{t=1}^T
%\left[ \frac{1}{M}\sum_{m=1}^M \tilde{w}^m_t\right] \left[\sum_{m=1}^M
%\omega^m_t\right] \\
%&=& \prod_{t=1}^T \left[ \frac{1}{M}\sum_{m=1}^M \frac{w^m_t}{\sum_{m=1}^M
%    \omega^m_t}\right] \left[\sum_{m=1}^M \omega^m_t\right] \\
%&=& \prod_{t=1}^T \frac{1}{M}\sum_{m=1}^M w^m_t,
%\end{eqnarray}
%\begin{equation}
%p(\mathbf{y}_{1:T}|\mathbf{x}_0,\boldsymbol{\theta}) \approx \prod_{t=1}^T
%\frac{1}{M}\sum_{m=1}^M w^m_t,
%\end{equation}

More rigorously, $p_1(\cdot)$ is a marginal of a distribution over the
extended space in which the particle filter operates, a space that includes
variables associated with the resampling mechanism~\citep[c.f. Equation 22
  of][]{Andrieu2010}:
\begin{equation}\label{eqn:psi}
\psi(\mathbf{u}_{1:T}^{1:M},a_{1:T}^{1:M}|\mathbf{x}_0,\boldsymbol{\theta}) =
\prod_{t=1}^T \left[ r(a_t^{1:M}|w_{t-1}^{1:M}) \prod_{m=1}^M
  p(\mathbf{u}_t^m) \right].
\end{equation}
Recall that $a_t^m$ is the index of the particle at time $t - 1$ which is the
ancestor of particle $m$ at time $t$. The function $r(\cdot)$ gives the
probability of these. For some time $s \leq T$, let $b_s^m$ denote the index
of the particle at time $s$ which is the ancestor of particle $m$ at time $T$,
obtained by recursively tracing the ancestor indices backward through time,
starting at $b^m_T = m$, then $b^m_{T - 1} = a^m_T, b^m_{T - 2} = a^{b^m_{T -
    1}}_{T-1}$, and so forth. For some specific $m \sim
\mathcal{C}(w^{1:M}_T)$, a sample of $p_1(\cdot)$ is then given by the
marginal
$\{\mathbf{u}_1^{b_1^m},\ldots,\mathbf{u}_T^{b_T^m}\}$~\cite{Andrieu2010}.

PMMH chains can be ``sticky'' if the likelihood estimates from
(\ref{eqn:likelihood}) are highly variable. A chain moving into a particular
state on the basis of an unusually large likelihood estimate tends to remain
there for a prolonged period before accepting a new proposal. The source of
variability is both sampling and resampling error in the particle
filter~\citep{Pitt2002,Lee2008}. The methods proposed in this work target a
reduction of the former for any fixed number of particles. Furthermore, the
likelihood estimates are heteroskedastic with respect to the parameters. For
fixed $M$, the stickiness of a PMMH chain varies across the space of
parameters, and from some regions a chain may not move to the vicinity of the
posterior distribution in reasonable time. The problem is particularly acute
when the process model informs the APF's importance proposals, as in all of
the strategies presented in \S\ref{sec:pf}. The effectiveness of the
importance proposals is then a function of the likelihood of, and behaviours
induced by, the current setting of the process model parameters. For example,
small process noise variance parameters will produce narrow proposal
distributions that may amplify weight variance, and so the variability of
likelihood estimates. The mixing properties of the model may also be affected
by gradient- and decay-related parameters.

\subsection{Assessing the mixing of PMMH}\label{sec:car}

To empirically explore the stickiness of PMMH chains under different particle
filtering strategies, consider each point $(\mathbf{x}_0,\boldsymbol{\theta})$
of a grid or set of points, and run some particle filtering method to be
assessed $L$ times on each of those points. Each run $i$ can be interpreted as
a sample from
$\psi(\mathbf{u}_{1:T}^{1:M},a_{1:T}^{1:M}|\mathbf{x}_0,\boldsymbol{\theta})$
in (\ref{eqn:psi}), and a marginal log-likelihood estimate $\hat{l}^i$
obtained (by using (\ref{eqn:likelihood}) and taking the logarithm). This
gives $L$ log-likelihood estimates $\hat{l}^1,\ldots,\hat{l}^L$.

\begin{figure}[tp]
\centering
\includegraphics[width=\textwidth]{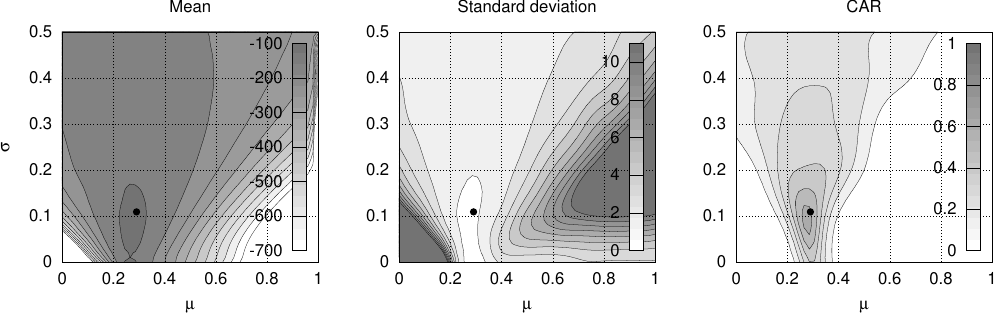}
\caption{Surfaces of \textbf{(a)} the mean of log-likelihood estimates,
  \textbf{(b)} the standard deviation of log-likelihood estimates, and
  \textbf{(c)} the CAR, for a bootstrap particle filter on the PZ case study
  of \S\ref{sec:pz}. The CAR surface gives a good idea of how variability in
  the likelihood estimator impacts the acceptance rate of a PMMH sampler
  according to its current state.}
\label{fig:pz_moments}
\end{figure}

At this point it is possible to compute a Monte Carlo estimate of the
log-likelihood:
\begin{equation}
\mathbb{E}(l) \approx \bar{l} = \frac{1}{L}\sum_{i=1}^L \hat{l}^i.
\end{equation}
and its standard deviation:
\begin{equation}
\sqrt{\mathbb{E}\left((l - \bar{l})^2\right)} \approx \sqrt{\frac{1}{L}\sum_{i=1}^L (\hat{l}^i - \bar{l})^2}.
\end{equation}
This approach is taken in \citet{Pitt2012} for a single central point of the
posterior distribution. The idea is readily extended across a grid or set of
values. Figure \ref{fig:pz_moments}(a~\&~b) do so, using the PZ model
considered in \S\ref{sec:pz}. The model has two parameters, $\mu$ and
$\sigma$. Estimates are made at each point of a 32 by 32 grid across the
support of the uniform prior distribution over parameters. The surface is then
interpolated using a Gaussian process fit by maximum likelihood, with a
constant mean function, isotropic squared exponential covariance function and
Gaussian likelihood \citep{Rasmussen2006,GPML}. \citet{Pitt2012} provides
guidance to set the number of particles according to standard deviation
estimates such as these.

We propose an alternative to standard deviation, which we call the
conditional acceptance rate (CAR). The intuition is to approximate, at
all points $(\mathbf{x}_0,\boldsymbol{\theta})$ of a grid or set, the
acceptance rate of a PMMH chain that starts at that point and remains there
indefinitely by using a Dirac $\delta$-function proposal centred at that
point. This can be seen as the limit of the acceptance rate when shrinking the
proposal distribution. For conventional MH with an exact likelihood this will
always be one, but for methods using a likelihood estimator, such as PMMH,
this will be less than that owing to variance in the estimator. The CAR is
always a number on $(0,1]$. While related to the standard deviation, the CAR
  is more directly interpretable as to the impact of variability in the
  likelihood estimator on the acceptance rate of a chain, and accommodates
  asymmetry in that variability. We prefer it for these reasons.

Consider a MH chain targeting $\psi(\mathbf{u}_{1:T}^{1:M},a_{1:T}^{1:M}|\mathbf{x}_0,\boldsymbol{\theta})$,
with independent proposal also
$\psi(\mathbf{u}_{1:T}^{1:M},a_{1:T}^{1:M}|\mathbf{x}_0,\boldsymbol{\theta})$,
but restricted to the $L$ discrete states already drawn from it. The
transition probability matrix $\mathbf{T} \in \mathbb{M}^{L \times L}$ of the
chain is:
\begin{equation}
T_{ij} = \begin{cases}
\frac{1}{L}\min\left[\exp(\hat{l}^j - \hat{l}^i), 1\right] & \quad
\text{if $i \neq j$} \\
1 - \sum_{\substack{k = 1 \\ k \neq i}}^L T_{ik} & \quad \text{if $i = j$}.
\end{cases}
\end{equation}
$T_{ij}$ gives the probability of moving to the $j$th state, of log-likelihood
$\hat{l}^j$, from the $i$th state, of log-likelihood $\hat{l}^i$. From state
$i$, the probability of an acceptance occurring in the next step, marginalised
over all possible proposals, is
\begin{equation}
\beta^i = 1 - T_{ii} + 1/L.
\end{equation}
The $1/L$ bias is incurred by using only a finite number of likelihood
estimates.

From some arbitrary state, the Markov model defined by $\mathbf{T}$ may be run
to equilibrium, where the probability of being in state $i$ is simply the
normalised term
\begin{equation}\label{eqn:p}
p^i = \frac{\exp \hat{l}^i}{\sum_{j=1}^L \exp \hat{l}^j}.
\end{equation}
The long-term acceptance rate, which we refer to as the conditional acceptance
rate at a given point, is then
\begin{equation}\label{eqn:alpha}
CAR(\mathbf{x}_0,\boldsymbol{\theta}) = \sum_{i=1}^L p^i \beta^i.
\end{equation}
Note that if all log-likelihood estimates are the same at a point, then
$CAR(\mathbf{x}_0,\boldsymbol{\theta}) = 1$. This is most easily seen through
(\ref{eqn:alpha}), as in this case $\beta^i = 1$ for all $i = 1,\ldots,L$, and
the remaining sum over $p^i$ is necessarily 1. Given that a finite number of
log-likelihood estimates are used, in the worst case
$CAR(\mathbf{x}_0,\boldsymbol{\theta})$ is still greater than $1/L$.

Figure \ref{fig:pz_moments}(c) depicts the CAR surface computed across the
same grid and same log-likelihood estimates as preceding plots in the same
figure. This gives a clear picture that mixing is best in the high-likelihood
region, declining with anisotropy away from that region.

In practice, the computation of CAR is simplified by following the procedure
in Appendix \ref{app:car}.

%%%%
\section{Case studies in marine biogeochemistry}\label{sec:cases}

The proposed methods are assessed empirically on two models in the domain of
marine biogeochemistry. All methods are assessed in each of two
configurations, as in Table \ref{tab:configs}. The first is
\emph{particle-matched}, where the number of particles, $M$, is the same for
all methods. The second is \emph{compute-matched}, where $M$ is adjusted for
all methods so as to roughly equate execution times. The latter is achieved by
matching the total number of propagations, where these include lookahead
pilots and UKF $\sigma$-points. We justify this by noting that for ordinary
differential equation models such as those considered here, propagation
typically dominates execution time (80-90\% in these cases). By controlling
the number of propagations, we roughly equate execution times in a fashion
that is independent of any particular implementation in code.

Simulated data is used, generated from the case study models themselves. This
has a number of advantages over real observational data: execution time can be
managed to facilitate the many runs required for some diagnostics, the model
is perfect in the generative sense, capturing all, and only, those processes
influencing observations, and a known ground truth for all latent variables is
available for validation of the methods. The second model is representative of
real-world usage, however, and is fit to observational data using simpler PMMH
methods in \citet{Parslow2013}.

%As the same model is used both for generating the data and doing inference, we
%refer to these as \emph{twin experiments}, a term borrowed from the data
%assimilation community.

%An observational data set is available, specifically from Ocean Station P
%(OSP), to which the second model has previously been fit using similar, but
%simpler, methods \citep{Parslow2013}. While the methods presented here can be
%used on this, the story for this data set is as yet incomplete, particularly
%around rigorous handling of its sparsity (up to seven weeks between
%observations, compared to daily for the simulated sets here). Until that story
%is complete, we have chosen not to draw on this data for testing purposes, but
%pick up on the point in discussion (\S\ref{sec:discussion}). Besides this
%issue of sparsity, however, we believe that the second case study is quite
%representative of real-world cases, especially in its nonlinear functional
%forms.

\begin{table}[tp]
\footnotesize
\centering
\begin{tabular}{|l|l|r|rr|rr|r|rr|rr|}
\toprule
& & \multicolumn{5}{c}{\emph{Particle-matched}} &
\multicolumn{5}{|c|}{\emph{Compute-matched}} \\
\hline
Case & Method & $M$ & \multicolumn{2}{r|}{Acceptance} &
\multicolumn{2}{r|}{ESS} & $M$ & \multicolumn{2}{r|}{Acceptance} &
\multicolumn{2}{r|}{ESS} \\
\midrule
PZ & PF0 & 64 & .182 & (.003) & 1444 & (129) & 384 & .245 & (.002) & 2021 & (196) \\
   & PF1 & 64 & .189 & (.003) & 1529 & (141) & 192 & .233 & (.002) & 1926 & (184) \\
   & MUPF0 & 64 & .208 & (.003) & 1707 & (156) & 376 & .251 & (.002) & 2080 & (200) \\
   & MUPF1 & 64 & .213 & (.003) & 1757 & (162) & 184 & .243 & (.002) & 2008 & (196) \\
   & CUPF0 & 64 & .209 & (.003) & 1709 & (155) & 64 & .209 & (.003) & 1709 & (155) \\
   & CUPF1 & 64 & .214 & (.003) & 1752 & (159) & 64 &  .214 & (.003) & 1752 & (159) \\
\midrule
NPZD & PF0 & 64 & .142 & (.003) & 115 & (8) & 1536 & .276 & (.002) & 163 & (16) \\
     & PF1 & 64 & .139 & (.006) & 110 & (9) & 768 & .254 & (.003) & 155 & (14) \\
     & MUPF0 & 64 & .165 & (.003) & 123 & (8) & 1504 & .278 & (.002) & 162 & (16) \\
& MUPF1 & 64 & .167 & (.007) & 121 & (10) & 736 & .263 & (.003) & 157 & (15) \\
     & CUPF0 & 64 & .169 & (.003) & 125 & (10) & 64 & .169 & (.003) & 125 & (10) \\
     & CUPF1 & 64 & .177 & (.005) & 125 & (9) & 64 & .177 & (.005) & 125 & (9) \\
\bottomrule
\end{tabular}

\caption{Configuration of methods for case studies, with mean (and standard
  deviation) of resulting acceptance rates and effective sample sizes across
  256 chains. Chains for the PZ case study are run for 50000 steps, and those
  for the NPZD case study for 75000 steps. In computing ESS, autocorrelations
  are truncated at a lag of 100 with the first 10000 steps removed for the PZ
  case study, and truncated at a lag of 400 with the first 25000 steps removed
  for the NPZD case study. In the particle-matched configurations,
  improvements in acceptance rates and ESS are clear for the UKF-based
  proposal strategies (MUPF and CUPF). In compute-matched configurations there
  is no clear improvement on these metrics over the bootstrap methods (PF).}
\label{tab:configs}
\end{table}

\subsection{PZ model}\label{sec:pz}

The first model considered is a variant of the Lotka-Volterra differential
system \citep{Lotka1925,Volterra1931}, specifically over the predator-prey
relationship of zooplankton and phytoplankton in a marine
environment. Previously treated with a PMMH-style sampler \citep{Jones2010},
the intent here is to plumb deeper into the behaviour of the algorithm, the
presence of just two parameters providing an ideal opportunity to visualise
dependence of CAR on $\boldsymbol{\Theta}$. This PZ (phytoplankton and
zooplankton) model modifies the classic Lotka-Volterra with the addition of a
quadratic mortality term for zooplankton and a stochastic growth term for
phytoplankton. The stochasticity admits varying growth rates in phytoplankton
without explicitly modelling contributory factors such as light and
temperature, and thus exemplifies how such uncertainties can be treated by the
introduction of stochasticity into an otherwise deterministic model
\citep{Jones2010,Parslow2013}.

The state of the model is given by $\mathbf{X} = \{P, Z, \alpha\}$,
with $P$ and $Z$ denoting concentrations of phytoplankton and zooplankton,
respectively, and $\alpha$ the stochastic growth rate of phytoplankton. These
interact via:
\begin{eqnarray}
\frac{dP}{dt} &=& \alpha_t P - cPZ\\
\frac{dZ}{dt} &=& ecPZ - m_lZ - m_q Z^2.
\end{eqnarray}
Here, $t$ is time in days, with prescribed constants $c = .25$, $e = .3$, $m_l
= .1$ and $m_q = .1$. While $P$ and $Z$ are modelled in continuous time, the
stochastic growth term, $\alpha_t$, is modelled in discrete time, updated
daily using $\alpha_t \sim \mathcal{N}(\mu,\sigma)$. Parameters to be
estimated are $\boldsymbol{\Theta} = \{\mu,\sigma\}$. Uniform prior
distributions are assigned to the parameters, $\mu \sim \mathcal{U}(0,1)$ and
$\sigma \sim \mathcal{U}(0,.5)$. Log-normal distributions are placed over the
initial conditions, $\ln P \sim \mathcal{N}(\ln 2, .2)$ and $\ln Z \sim
\mathcal{N}(\ln 2, .1)$. Phytoplankton ($P$) is observed with log-normal
noise:
\begin{equation}
\ln Y_P \sim \mathcal{N}(\ln P,.2).
\end{equation}

The differential equations must be numerically integrated forward in
time. While an Euler discretisation would yield a closed-form transition
density, the system is not numerically stable with such a low-order scheme. A
fourth-order scheme, precisely the low-storage Runge-Kutta method
RK4(3)5[2R+]C~\cite{Carpenter1994,Kennedy2000}, with adaptive time step, is
used~\cite{Murray2012}. This does not readily yield a closed-form transition
density, motivating the approach. Simulated data is used by integrating
forward a single trajectory for 100 days, taking $P$ daily and adding
observation noise.

The target is factorised as in (\ref{eqn:pi1}), so that initial conditions are
importance sampled within the particle filter. A systematic
resampler~\cite{Kitagawa1996} is used to minimise the contribution of the
resampler to the variance of the likelihood
estimator~\cite{Pitt2002,Lee2008}. To construct sensible starting and proposal
distributions for the MH chain through $\boldsymbol{\Theta}$, a joint UKF
(i.e. the state is augmented to include $\boldsymbol{\Theta}$) is first
applied. The final filtering distribution,
$\hat{p}(\boldsymbol{\theta}_T|\mathbf{y}_{1:T}) \equiv
\mathcal{N}(\hat{\boldsymbol{\mu}}_T, \hat{\Sigma}_T)$, is used as the
starting distribution for the MH chain, and its covariance, $\hat{\Sigma}_T$,
scaled by .18, for a random-walk Gaussian proposal. The scaling factor is
chosen using pilot runs with the PF0 method. Starting at the rule-of-thumb
$2.4^2/N_{\theta} = 2.88$ \citep{Gelman1994} (recall $\Theta \in
\mathbb{R}^{N_\theta}$), it is halved (four times) until a mixing rate close
to the rule-of-thumb 23\% \citep{Gelman1994} is achieved in the first 500
steps of the chain. Using this proposal, 256 PMMH chains of 50000 steps are
then run for each method.

\begin{figure}[tp]
\centering
\includegraphics[width=\textwidth]{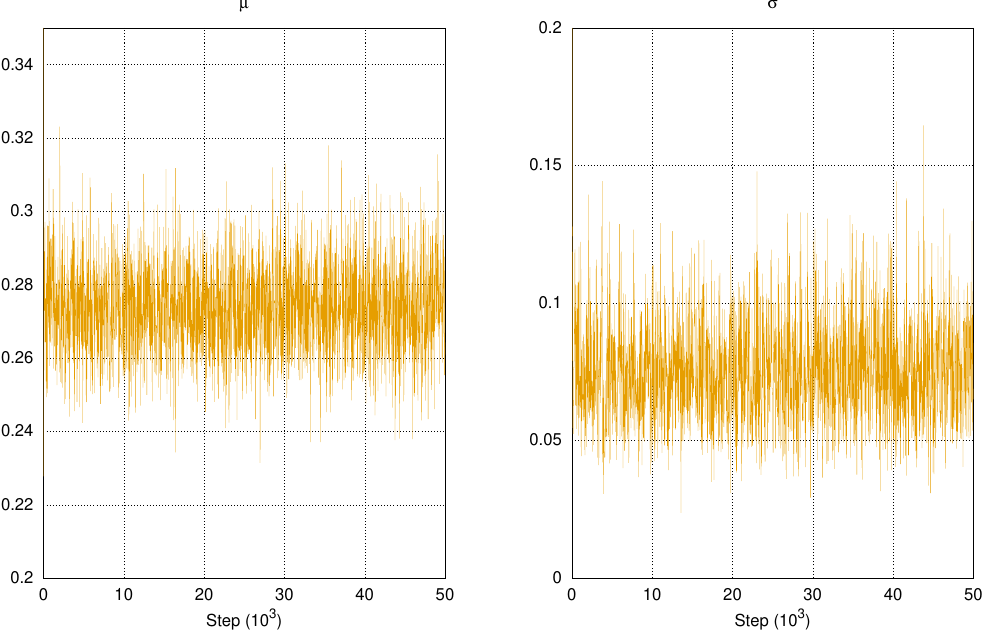}
\caption{Indicative trace plots of a single PMMH chain over parameters of the
  PZ model, using the PF0 method with 64 particles. These indicate good mixing
  of the chain. Other methods use the same proposal and achieve higher
  acceptance rates (Table \ref{tab:configs}); their trace plots, while not
  shown, appear at least as good on inspection.}
\label{fig:pz_traces}
\end{figure}

\begin{figure}[tp]
\centering
\includegraphics[width=\textwidth]{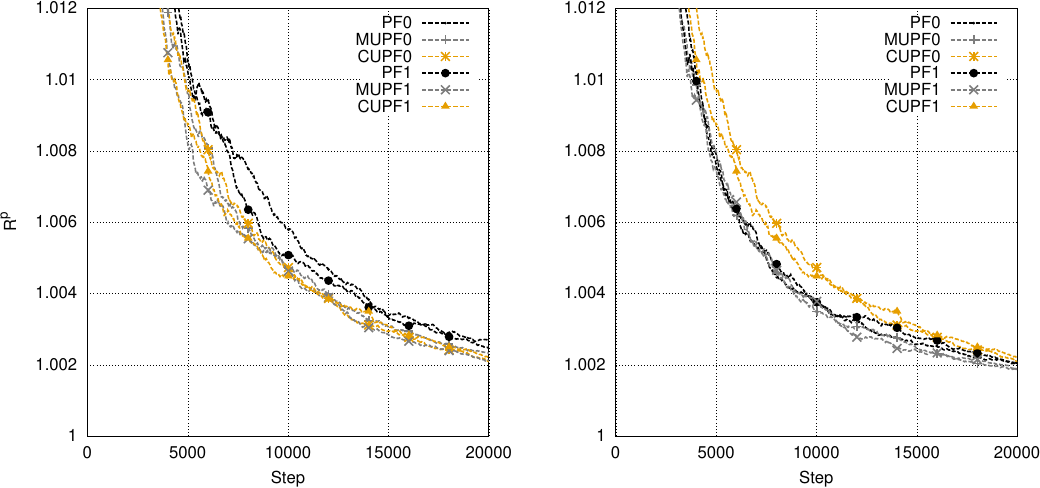}
\caption{Convergence rates of Markov chains for the PZ case study,
  \textbf{(a)} particle-matched, and \textbf{(b)} compute-matched. Each line
  shows the evolution of the $\hat{R}^p$ statistic of \citet{Brooks1998} for a
  particular method as the number of steps taken increases. The statistic is
  computed using 256 chains for each method. A more rapid approach to 1
  indicates faster convergence. In the particle-matched configuration the
  UKF-based methods demonstrate faster convergence. In the compute-matched
  configuration the MUPF methods may give a slight improvement over the
  simpler PF methods.}
\label{fig:pz_converge}
\end{figure}

Performance is first assessed with established metrics. Trace plots for a
single chain of the PF0 method with $M = 64$ particles are given in Figure
\ref{fig:pz_traces}. These indicate good mixing. The other methods, not shown,
produce traces that also indicate good mixing. Table \ref{tab:configs}
provides the mean and standard deviation of acceptance rates and effective
sample sizes (ESS) across all chains for each method. The ESS is computed
separately for each parameter of each chain~\cite{Kass1998}:
\begin{equation}
ESS = 1 + 2\sum_{k=1}^\infty R(k,\theta),
\end{equation}
where $R(k,\theta)$ is the lag-$k$ autocorrelation of the single parameter
$\theta$. The first 10000 steps are removed as burn-in and the infinite sum
truncated at $k = 100$. The minimum ESS across all parameters of a chain is
then taken as that chain's overall ESS for reporting in Table
\ref{tab:configs}. Finally, the multivariate $\hat{R}^p$ statistic of
\citet{Brooks1998} is computed across all chains to empirically assess the
rate of convergence (Figure \ref{fig:pz_converge}). All of these metrics
establish that the chains are mixing well.

\begin{figure}[tp]
\centering
\includegraphics[width=\textwidth]{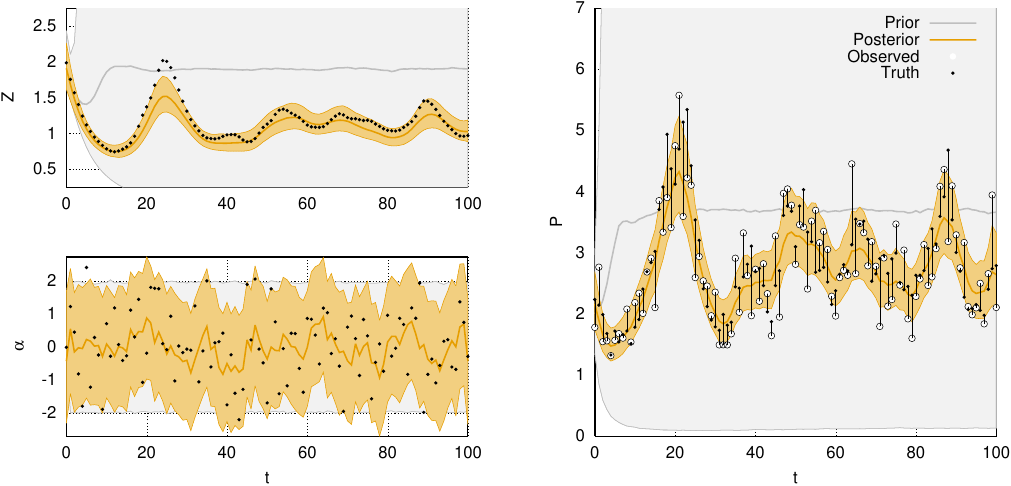}
\caption{Time-marginal posteriors over state variables ($P$ and $Z$) and noise
  term ($\alpha$) for the PZ case study. Results are obtained by PMMH using
  the CUPF1 method. Other methods give comparable results, although not
  shown. The bold centre lines of the prior and posterior distributions denote
  their medians, and the shaded regions their 95\% credibility intervals.}
\label{fig:pz_state}
\end{figure}

\begin{figure}[tp]
\centering
\includegraphics[width=\textwidth]{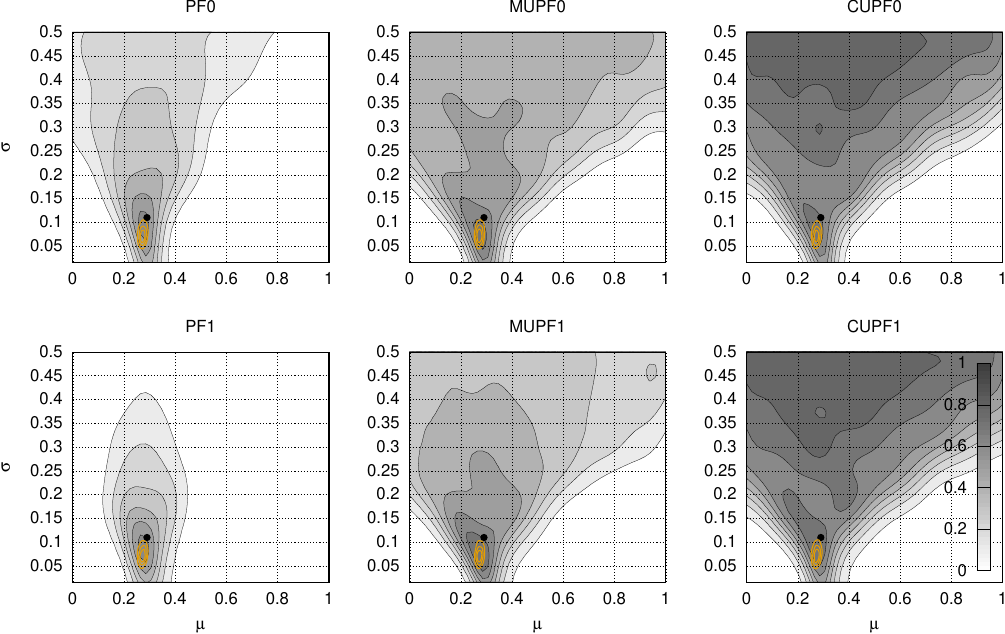}
\caption{CAR surfaces for the PZ case study across the support of the uniform
  prior distribution over parameters. Each plot shows the results for a
  particular method in its particle-matched configuration. Darker shading
  denotes higher CAR (see key bottom right). The dots at approximately (0.3,
  0.1) in each plot mark the ground truth parameters from which the data set
  is simulated. The bold contours nearby mark the posterior distribution
  obtained. A clear decline in CAR is evident as distance increases from the
  ground truth and posterior region. The CUPF methods appear more robust than
  others at high $\sigma$.}
\label{fig:pz_acceptance}
\end{figure}

\begin{figure}[tp]
\centering
\includegraphics[width=\textwidth]{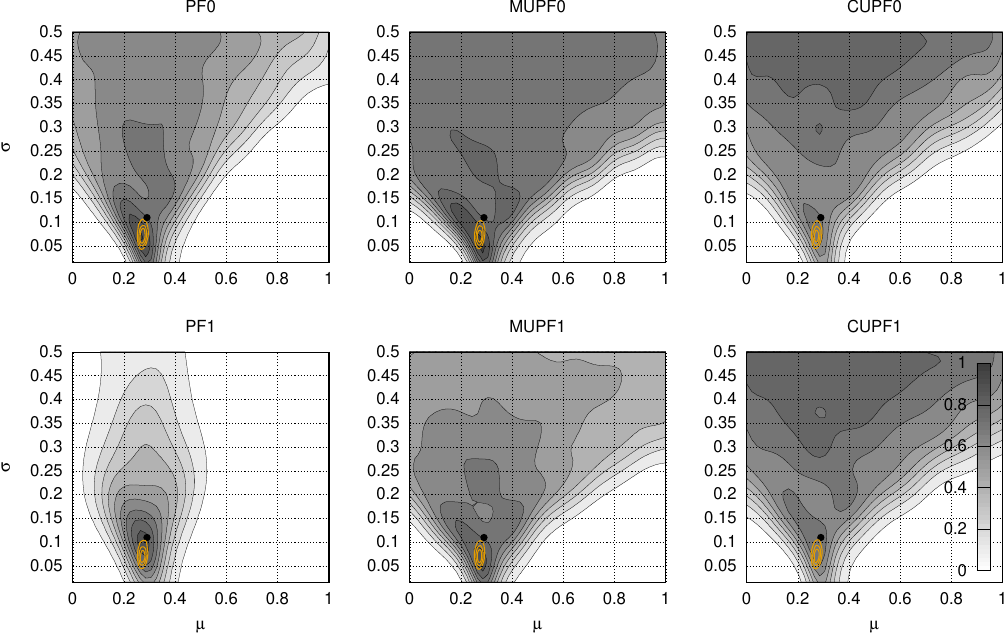}
\caption{CAR surfaces for the PZ case study for compute-matched
  configurations. See Figure \ref{fig:pz_acceptance} caption for details.}
\label{fig:pz_acceptance_pmatch}
\end{figure}

The posterior distribution obtained over parameters is marked in Figures
\ref{fig:pz_acceptance} and \ref{fig:pz_acceptance_pmatch}, using samples
drawn across all chains for each method. For one chain, the state posterior is
visualised in Figure \ref{fig:pz_state}, along with the ground truth
trajectory and observations for comparison.

With results looking sensible so far, we proceed with a comparison using the
CAR metric introduced in \S\ref{sec:car}. For each method, the CAR is computed
at 1024 points on a 32 by 32 regular grid across the uniform prior
distribution, using 200 likelihood evaluations at each point. To produce
contours of the surface, a Gaussian process is fit to the points by maximum
likelihood, with a constant mean function, isotropic squared exponential
covariance function and Gaussian likelihood
\citep{Rasmussen2006,GPML}. Results are presented for particle-matched
configurations in Figure \ref{fig:pz_acceptance}, and for compute-matched in
Figure \ref{fig:pz_acceptance_pmatch}.

\subsection{NPZD model}\label{sec:npzd}

The introduction of nutrients, $N$, and detritus, $D$, into the PZ model
provides a more realistic system with which real observational data can begin
to be assimilated: an NPZD model. These additional terms are accompanied by
various environmental forcings and rate processes that produce a more
challenging model, with nonlinear responses ranging from convergence, to
periodicity, to chaos. The full details and motivation behind the model are
given in \citet{Parslow2013}. A brief description to elucidate some of the
complexity is given here.

The NPZD model represents the interaction of nutrients ($N$), phytoplankton
($P$), zooplankton ($Z$) and detritus ($D$), quantified in the common currency
of nitrogen, within the surface mixed layer of a body of water. The surface
waters are modelled as a single box, subject to exogenous environmental
forcings such as available light, temperature and changes in mixed layer
depth.

%For reference as components of the model are introduced, the model state is
%$\mathbf{X} = \{P,$ $Z,$ $D,$ $N,$ $g^{\text{max}},$ $\lambda^{\text{max}},$
%$R_N,$ $a_N,$ $I_Z,$ $Cl_Z,$ $E_Z,$ $r_D,$ $m_Q\}$, and the parameters
%$\boldsymbol{\Theta} = \{K_W,$ $a_{Ch},$ $S_D,$ $f_D,$ $\mu_{g^{\text{max}}},$
%$\mu_{\lambda^{\text{max}}},$ $\mu_{R_N},$ $\mu_{a_N},$ $\mu_{I_Z},$
%$\mu_{Cl_Z},$ $\mu_{E_Z},$ $\mu_{r_D},$ $\mu_{m_Q},$ $PDF,$
%$ZDF\}$.

\subsubsection{Noise model}

The model features nine noise terms, $\xi_i$ for $i = 1,\ldots,9$, each
coupled to a univariate autoregressive process $B_i$. Four of these are
phytoplankton-related, given by
\begin{equation} 
B_i(t + \Delta t) = B_i(t) \cdot (1 - \Delta t/\tau_P) + (\mu_i +
PDF \cdot \sigma_i \xi_i) \cdot \Delta t/\tau_P,
\end{equation}
where $\Delta t$ is a discrete time step (one day), $\mu_i$ a parameter to be
estimated, $PDF$ a common \emph{diversity factor} parameter to be estimated,
$\sigma_i$ a prescribed scaling factor, and $\tau_P$ a common characteristic
time scale, also prescribed. The remaining five autoregressive processes are
zooplankton-related, modelled using the same form, with $ZDF$ and $\tau_Z$
replacing $PDF$ and $\tau_P$, respectively.

Each process represents a property of the phytoplankton (zooplankton)
community, the species composition of which will change with time. Rather than
model individual species, the phytoplankton (zooplankton) community is
modelled collectively, with diversity factors $PDF$ and $ZDF$ scaling
stochastic drivers used to model the changing influence of community
composition.

The four phytoplankton processes are $\{g^{\text{max}}, \lambda^{\text{max}},
R_N, a_N\}$, and the five zooplankton processes $\{I_Z, Cl_Z, E_Z, r_D,
m_Q\}$. Each is accompanied by its matching noise term amongst $\{
\xi_{g^{\text{max}}},$ $\xi_{\lambda^{\text{max}}},$ $\xi_{R_N},$ $\xi_{a_N},$
$\xi_{I_Z},$ $\xi_{Cl_Z},$ $\xi_{E_Z},$ $\xi_{r_D},$ $\xi_{m_Q}\}$, and mean
parameter amongst $\{ \mu_{g^{\text{max}}},$ $\mu_{\lambda^{\text{max}}},$
$\mu_{R_N},$ $\mu_{a_N},$ $\mu_{I_Z},$ $\mu_{Cl_Z},$ $\mu_{E_Z},$ $\mu_{r_D},$
$\mu_{m_Q}\}$.

\subsubsection{Process model}

The remaining state variables are $\{N,P,Z,D\}$ and parameters $\{K_W,$
$a_{Ch},$ $S_D,$ $f_D\}$. The equations governing interactions between the
remaining state variables are:
\begin{eqnarray}
\frac{dN}{dt} &=& -g\cdot P + (1 - E_Z)\cdot (1- f_D)\cdot gr\cdot Z + r\cdot
    D + \frac{\kappa}{MLD} \cdot (BCN - N) \\
\frac{dP}{dt} &=& g\cdot P-gr\cdot Z + \frac{\kappa}{MLD} \cdot
(BCP-P) \\
\frac{dZ}{dt} &=& E_Z\cdot gr\cdot Z - m\cdot Z \\
\frac{dD}{dt} &=& (1 - E_Z)\cdot f_D\cdot gr\cdot Z + m\cdot Z - r\cdot D -
    S_D\cdot \frac{D}{MLD} + \frac{\kappa}{MLD} \cdot (BCD-D).
\end{eqnarray}
Here, $g$ is the phytoplankton specific growth rate (per day, or d$^{-1}$),
$gr$ is the zooplankton specific grazing rate (mg $P$ grazed per mg $Z$
d$^{-1}$), $m$ is the zooplankton specific mortality rate (d$^{-1}$), and $r$
is the specific breakdown rate of detritus (d$^{-1}$). A fraction, $E_Z$, of
zooplankton ingestion is converted to zooplankton growth and, of the
remainder, a fraction, $f_D$, allocated to detritus and the rest released as
dissolved inorganic nutrient, $N$.

The rate processes $gr$, $m$ and $g$ are not only functions of the state, but
also prescribed exogenous forcings and physiological constants. A
multiplicative temperature correction $Tc$ is applied to all of these, for
which a $Q_{10}$ formulation for dependence on temperature, $T$, is used:
\begin{equation}
Tc = Q_{10}^{(T - T_{\text{ref}})/10},
\end{equation}
where $T_{\text{ref}}$ is a reference temperature, and $Q_{10}$ a prescribed
constant.

The zooplankton grazing rate, $gr$, is dependent on the relative availability
of phytoplankton, $A$:
\begin{equation}
  gr = \frac{Tc\cdot I_Z\cdot A^\upsilon}{(1 + A^\upsilon )}, \label{eq:gr}
\end{equation}
where $\upsilon$ is a given power, and
\begin{equation}
  A = \frac{Cl_Z\cdot P}{I_Z}.
\end{equation}
$I_Z$ is the maximum zooplankton ingestion rate (mg $P$ per mg $Z$ per day);
$Cl_Z$ is the maximum clearance rate (volume in m$^3$ swept clear per mg $Z$
per day).  For $\upsilon = 1$, (\ref{eq:gr}) takes the form of a Type-2
functional response (standard rectangular hyperbola) \citep{Holling1966}, and
for $\upsilon >1$ a Type-3 sigmoid functional response.

A quadratic formulation for zooplankton mortality is adopted after
\citet{Steele1976} and \citet{Steele1992}:
\begin{equation}
m = Tc\cdot m_Q\cdot Z,
\end{equation}
where the quadratic mortality rate, $m_Q$, has units of d$^{-1}(\text{mg} Z
\text{m}^{-3})^{-1}$. The detrital remineralisation rate is dependent only on
temperature:
\begin{equation}
r = Tc\cdot r_D,
\end{equation}
where $r_D$ prescribes the remineralisation rate at a reference temperature.

The phytoplankton specific growth rate, $g$, depends on temperature, $T$,
available light or irradiance, $E$, and dissolved inorganic nutrient, $N$. It
is expressed in terms of a maximum specific growth rate at the reference
temperature, $g^{\text{max}}$ (d$^{-1}$), a light-limitation factor, $h_E$, and
a nutrient-limitation factor, $h_N$:
\begin{equation}
g = Tc\cdot g^{\text{max}}\cdot h_E\cdot h_N/(h_E + h_N).
\end{equation}
The light-limitation factor is given by
\begin{equation}
h_E = 1 - \exp(-\alpha\cdot \lambda^{\text{max}}\cdot E/g^{\text{max}}),
\end{equation}
where $\alpha$ is the initial slope of the photosynthesis versus irradiance
curve (mg $C$ mg $Chla^{-1}$ mol photon$^{-1}$ m$^2$), and
$\lambda^{\text{max}}$ is the maximum $Chla:C$ (chlorophyll-a to carbon) ratio
(mg $Chla$ mg $C^{-1}$). Here, $\alpha$ is calculated as the product of the
chlorophyll-specific absorption coefficient for phytoplankton, $a_{Ch}$ (m$^2$
mg $Chla^{-1}$), and the maximum quantum yield for photosynthesis, $Q$ (mg $C$
mol photons$^{-1}$). $E$ is the mean photosynthetic available radiation (PAR)
in the mixed layer and is given by
\begin{equation}
E = E_0 \cdot (1 - \exp(-Kz))/Kz,
\end{equation}
where $E_0$ is the mean daily photosynthetically available radiation (PAR) just below the air-sea interface, $Kz$ is given by
\begin{equation}
Kz = (K_W + a_{Ch} \cdot Chla) \cdot MLD.\label{eq:az}
\end{equation}
and $K_W$ is attenuation due to the seawater and $a_{Ch}$.

The nutrient-limitation factor is given by
\begin{equation}
h_N = \frac{N}{(g^{\text{max}} \cdot Tc/a_N) + N},
\end{equation}
where $a_N$ is the maximum specific affinity for nitrogen uptake
(d$^{-1}$ mg $N^{-1}$ m$^3$).

The phytoplankton $N:C$ (nitrogen to carbon) ratio, $\chi$, predicted by the
model is given by
\begin{equation}
\chi = \frac{\chi^{\text{min}}\cdot h_E + \chi^{\text{max}}\cdot h_N} {h_E + h_N},
\end{equation}
where $\chi^{\text{min}}$ and $\chi^{\text{max}}$ are the prescribed minimum
and maximum $N:C$ ratios (mg $N$ mg $C^{-1}$).

\subsubsection{Boundary conditions}

The simple single-box mixed layer model adopted here needs to allow for the
effects of physical exchanges between the mixed layer and the underlying water
mass. With the exception of $BCN$, all boundary conditions ($BCP,BCD,BCZ$) are
set to zero for the experiments in this work. The variable $\kappa$ sets the
strength of the mixing; in this study, we assume that the lower two metres of
the mixed layer are replenished daily with water from below. $MLD$ and $BCN$
are in this case time-invariant, and set to 40 m and 200 mg N m$^3$,
respectively.

\subsubsection{Observation model}

The model predicts the phytoplankton $Chla:C$ ratio $\lambda$, and
this can be combined with the $N:C$ ratio $\chi$ to convert phytoplankton
biomass $P$ (mg $N$ m$^{-3}$) to a predicted $Chla$ concentration:
\begin{equation}
Chla = P\cdot (\lambda^{\text{max}}/\chi^{\text{max}})\cdot h_N \cdot Tc /(R_N \cdot h_E + h_N).
\end{equation}

Both $N$ and $Chla$ are observed, each with log-normal noise of 40\%,
i.e. $\ln Y_N \sim \mathcal{N}(\ln N, .4)$, and $\ln Y_{Chla} \sim
\mathcal{N}(\ln Chla, .4)$. Observations are thus written $\mathbf{Y} = \{Y_N,
Y_{Chla}\}$.

\subsubsection{Experiments}

The fourth order Runge-Kutta scheme RK4(3)5[2R+]C~\cite{Kennedy2000} is again
used to numerically integrate the differential equations forward. Use of such
a higher-order scheme is essential for this model, which is unstable under
low-order schemes like Euler. A data set is generated by simulating the model
with artificial forcing for 100 days, from which nutrient ($N$) and
chlorophyll-a ($Chla$) observations are produced daily. A systematic
resampler~\cite{Kitagawa1996} is again used.

\begin{figure}[tp]
\centering
\includegraphics[width=\textwidth]{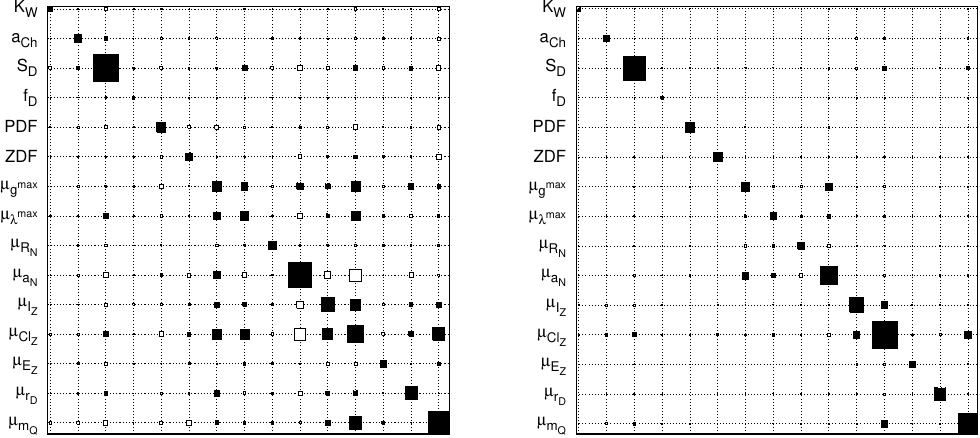}
\caption{Covariance matrices over parameters of the NPZD model, as returned by
  \textbf{(a)} PMMH using CUPF1, and \textbf{(b)} URTSS. The area of each
  square is proportional to magnitude, with filled squares denoting positive,
  and empty squares negative, covariance. The covariance matrices for initial
  conditions and cross-covariance (neither shown) also show similarity,
  although there is little interesting off-diagonal structure. While
  approximate, the URTSS method is inexpensive, and still captures some
  significant off-diagonal elements that can be used to construct a good
  proposal distribution for PMMH.}
\label{fig:npzd_paramcov}
\end{figure}

The target is factorised according to (\ref{eqn:pi2}), so that both parameters
and initial conditions are sampled by the MH chain. To construct sensible
starting and proposal distributions, a joint UKF is first applied, in the same
way as for the PZ model. Because starting and proposal distributions over both
parameters and initial conditions are now required, we then apply a joint
unscented Rauch-Tung-Striebel smoother (URTSS) \citep{Sarkka2008} to the
output of the UKF, giving a Gaussian approximation to the smoothing
distribution $p(\mathbf{x}_0,\boldsymbol{\theta}_0|\mathbf{y}_{1:T})$. This is
taken as the starting distribution for the MH chain, and its covariance,
scaled by .012, for a random-walk Gaussian proposal. Figure
\ref{fig:npzd_paramcov} shows a comparison of the covariance matrix obtained
by URTSS to that eventually obtained by PMMH; the similarity makes clear the
utility of the approach in constructing a sensible proposal
distribution. Using this proposal, 256 PMMH chains of 75000 steps are then run
for each method.

\begin{figure}[tp]
\centering
\includegraphics[width=0.8\textwidth]{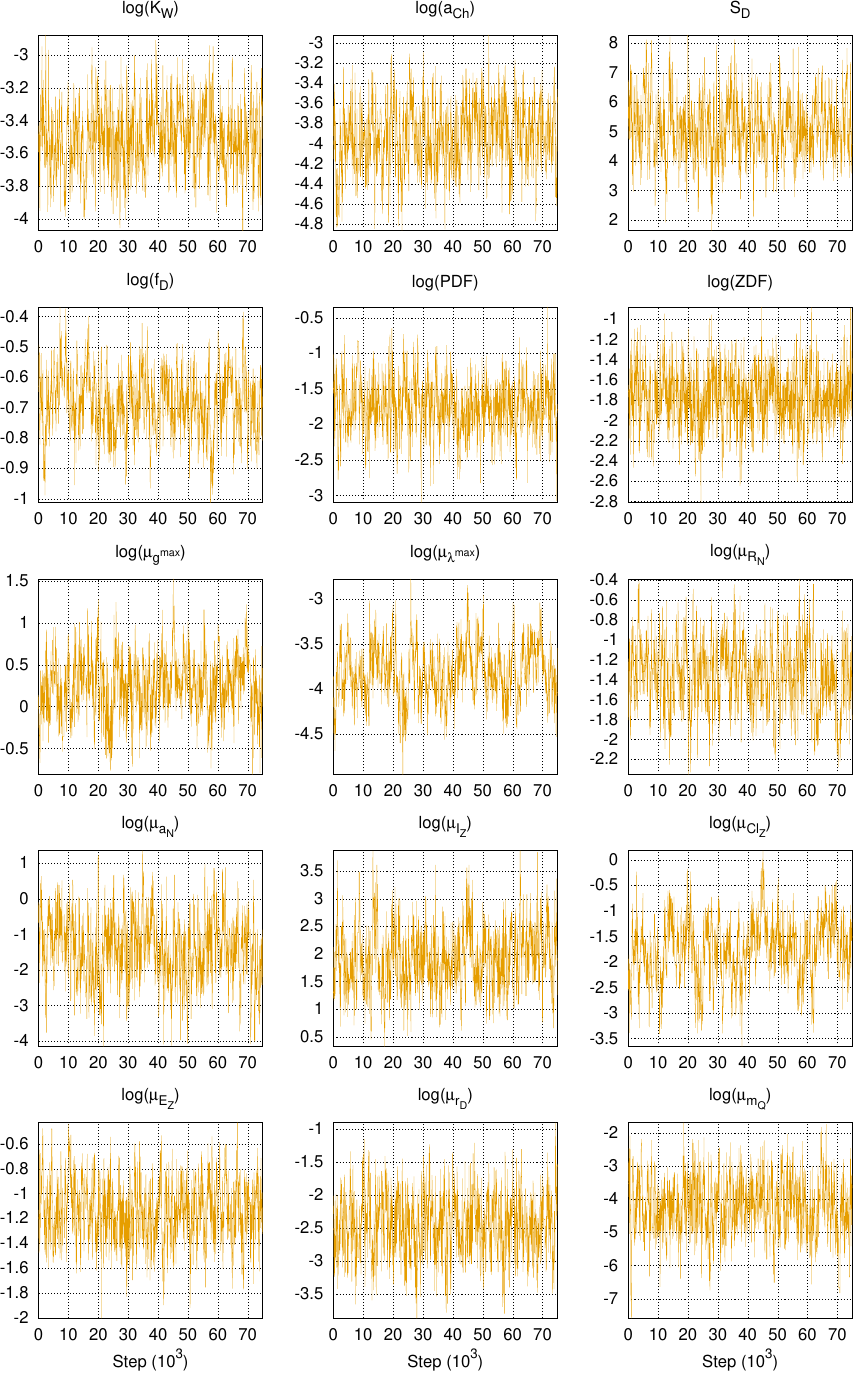}
\caption{Indicative trace plots of a single PMMH chain over parameters of the
  NPZD model, using the PF0 method with 64 particles. These indicate
  reasonable mixing of the chain. Some autocorrelation is apparent, reducing
  ESS in Table \ref{tab:configs}.}
\label{fig:npzd_traces}
\end{figure}

\begin{figure}[tp]
\centering
\includegraphics[width=\textwidth]{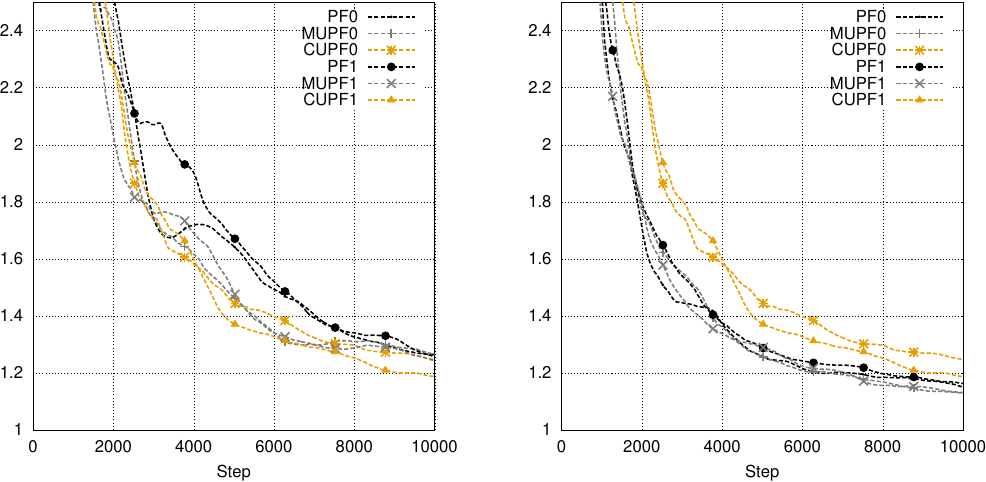}
\caption{Convergence rates of Markov chains for the NPZD case study,
  \textbf{(a)} particle-matched, and \textbf{(b)} compute-matched. See Figure
  \ref{fig:pz_converge} caption for details. In particle-matched
  configurations the UKF-based methods offer a clear improvement, but in
  compute-matched configurations the simpler PF0 and PF1 methods are again
  competitive.}
\label{fig:npzd_converge}
\begin{minipage}{\textwidth}
\end{minipage}
\end{figure}

\begin{figure}[tp]
\centering
\includegraphics[width=\textwidth]{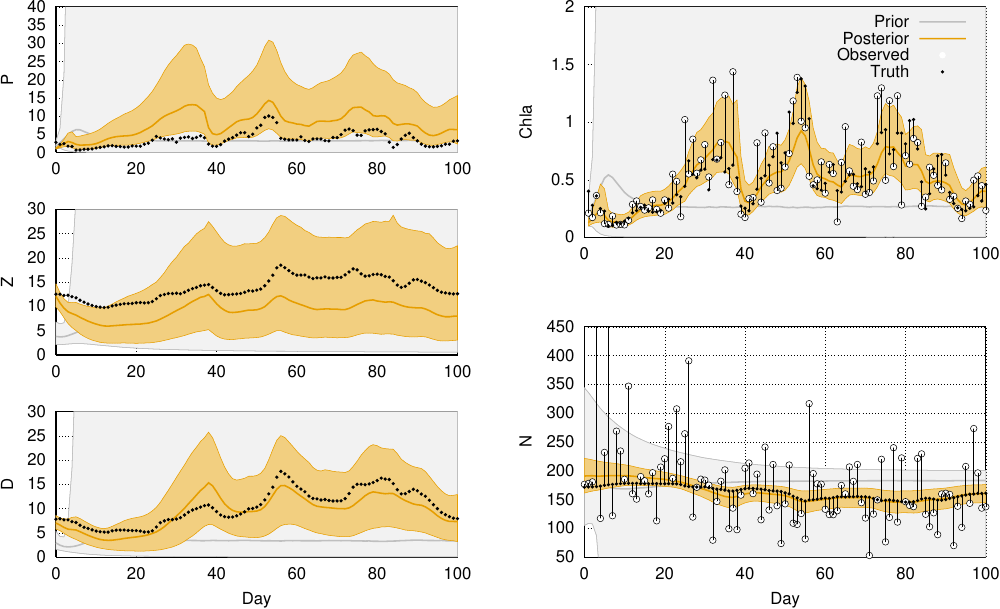}
\caption{Time-marginal posterior distributions over state variables for the
  NPZD case study. Results are obtained by PMMH using the CUPF1
  method. Other methods give comparable results, although not shown. The bold
  centre lines of the prior and posterior distributions denote their medians,
  and the shaded regions their 95\% credibility intervals.}
\label{fig:npzd_state}
\end{figure}

\begin{figure}[p]
\centering
\includegraphics[width=0.8\textwidth]{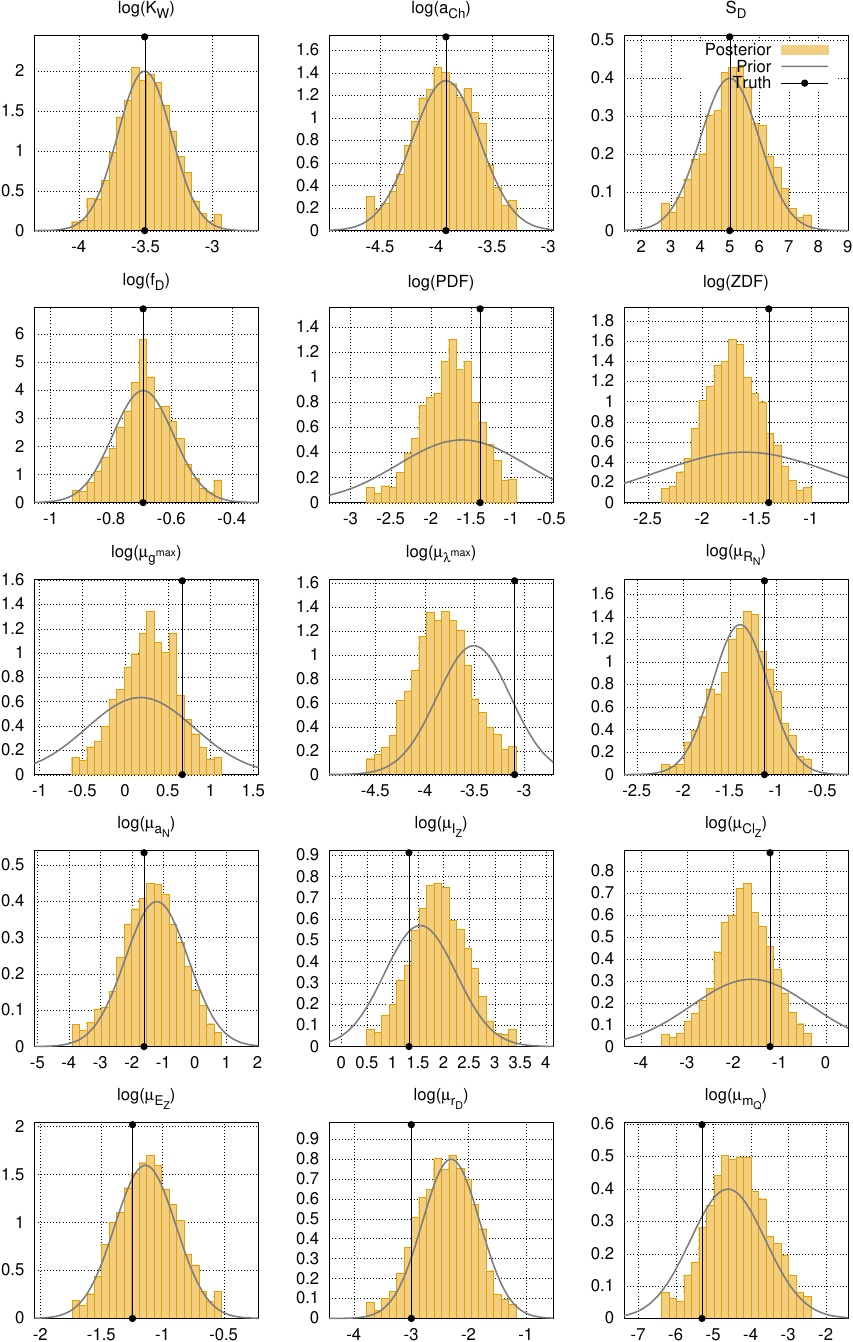}
\caption{Prior and posterior distributions over parameters for the NPZD case
  study. Results are obtained by PMMH using the CUPF1 method.}
\label{fig:npzd_parameters}
\end{figure}

Trace plots for a single chain of the PF0 method are given in Figure
\ref{fig:npzd_traces}. Table \ref{tab:configs} provides the mean and standard
deviation of acceptance rates and ESS across all chains for each method. In
computing ESS, the first 25000 samples from each chain are removed as burn-in,
and the infinite sum truncated at a lag of $k = 400$. The $\hat{R}^p$
statistic \citep{Brooks1998} is computed across multiple chains and shown in
Figure \ref{fig:npzd_converge}. All of these metrics indicate reasonable
mixing, albeit with effective sample size significantly lower than in the PZ
case study. The univariate posterior marginal distributions of parameters and
state are given in Figures \ref{fig:npzd_state} and \ref{fig:npzd_parameters}.

\begin{figure}[tp]
\centering
\includegraphics[width=\textwidth]{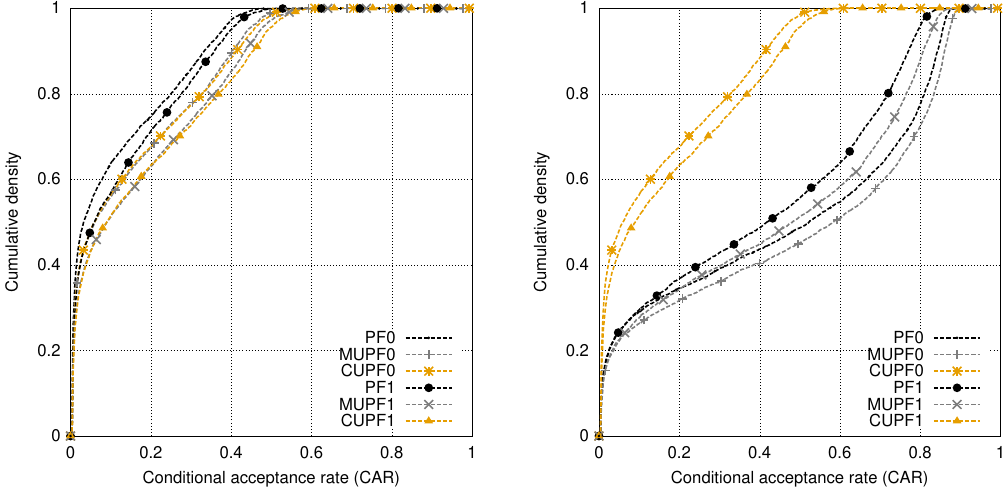}
\caption{Empirical cumulative distribution functions of the CAR for each
  method on the NPZD case study, \textbf{(a)} particle-matched, and
  \textbf{(b)} compute-matched. For each method, the CAR is computed at the
  same sample points used to construct Figure \ref{fig:npzd_copulas}. The
  empirical cumulative distribution function over all of these CARs is then
  evaluated. As higher CARs are preferred, a lower cumulative density on the
  $y$-axis is preferred for any given point on the $x$-axis. An advantage for
  the UKF-based methods is apparent in the particle-matched case, but this is
  only maintained for the MUPF0 method in the compute-matched case.}
\label{fig:npzd_acceptcdf}
\end{figure}

\begin{figure}[p]
\centering
\includegraphics[width=\textwidth]{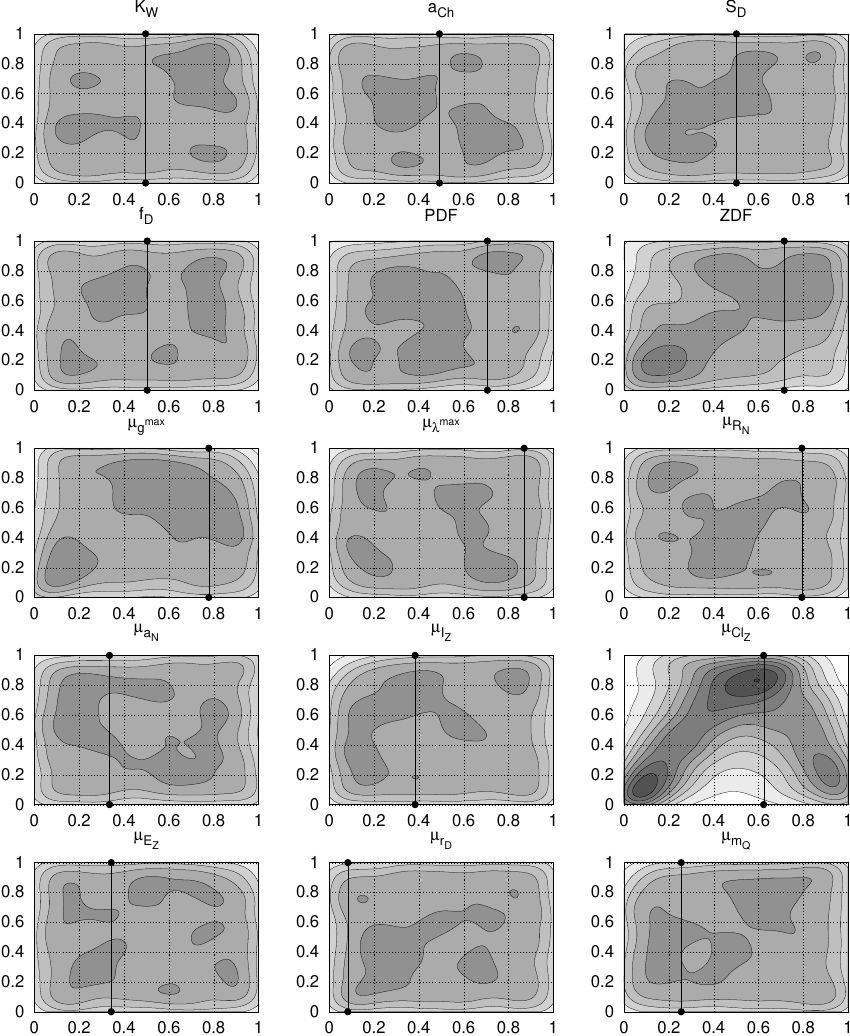}
\caption{Estimated copula functions between parameters ($x$-axes) and CAR
  ($y$-axes) for the NPZD case study, using the CUPF1 method. Other methods
  give similar results. For parameters, the prior univariate cumulative
  density functions are used for transformation to uniform marginals.  For the
  CAR, the empirical cumulative density function is used. The copula function
  is approximated using a kernel density estimate of bandwidth .075 over the
  4096 points sampled from the prior distribution over parameters, with the
  CAR computed at each point using 200 likelihood evaluations. Edge effects
  are an artifact of the kernel density estimate. Most striking is the
  significant sensitivity of the CAR to the $\mu_{Cl_Z}$ parameter, explained
  in the text.}
\label{fig:npzd_copulas}
\end{figure}

CARs are computed at a set of 4096 points drawn randomly from the prior
distribution. Because of the higher dimensionality of the NPZD model, plots
such as Figures \ref{fig:pz_acceptance} and \ref{fig:pz_acceptance_pmatch}
produced for the PZ model are not feasible. Instead, pairwise copulas between
parameters and the CAR, computed empirically, are shown in Figure
\ref{fig:npzd_copulas}, with the empirical cumulative distribution function of
the same CARs shown in Figure \ref{fig:npzd_acceptcdf}.

%%%%
\section{Discussion}\label{sec:discussion}

When the number of particles is matched across methods, the MUPF and CUPF
methods outperform the basic PF methods for both the PZ and NPZD cases: in
acceptance rate and ESS (Table \ref{tab:configs}), convergence rates (Figures
\ref{fig:pz_converge}(a) and \ref{fig:npzd_converge}(a)) and CAR (Figures
\ref{fig:pz_acceptance} and \ref{fig:npzd_acceptcdf}(a)). For the PZ model,
the lookahead degrades performance for PF1 and MUPF1, but not for CUPF1. This
is presumably because the single-point pilots used in the first two methods
are not representative of the whole predictive distribution. For the NPZD model,
the lookahead is beneficial. This is attributed to the NPZD model having
longer memory than the faster-mixing PZ model, a scenario where lookaheads
tend to be more useful.

In compute-matched configurations, the MUPF methods appear to retain some
advantage in the PZ case study: in acceptance rate and ESS (Table
\ref{tab:configs}), convergence rates (Figure \ref{fig:pz_converge}(b)) and
CAR (Figure \ref{fig:pz_acceptance_pmatch}). In the NPZD case study, overall
acceptance rates and ESS are very similar to the simpler PF methods (Table
\ref{tab:configs}), although there is some suggestion that at least the MUPF0
method retains an advantage in CAR across the space of parameters (Figure
\ref{fig:npzd_acceptcdf}(b)). The CUPF methods appear to offer no overall
advantage in compute-matched configurations, but an interesting subplot arises
from the CUPF methods in the PZ case study: no other methods match the CARs
achieved by them in low-likelihood regions, even after correction for compute
time (Figures \ref{fig:pz_acceptance} and
\ref{fig:pz_acceptance_pmatch}). This suggests that these methods may make a
more robust choice for early steps in a PMMH chain if a good initialisation is
unavailable, later displaced by one of the cheaper methods once in a region of
higher likelihood.

The MUPF method performs significantly better on the PZ model than the NPZD
model, and the PZ model is known to mix faster than the NPZD model. This is
consistent with the expectation that the MUPF methods should work better for
faster mixing models. An outstanding challenge is to design more generally
applicable proposal schemes for the disturbance state-space model that are
computationally competitive, and deliver more convincing outcomes for harder
cases such as the NPZD model. This is left to future work. There is, however,
evidence here that there exist proposals, enabled by the disturbance
state-space model formulation, that can improve PMMH performance in some
circumstances.

%In all methods, the overall acceptance rate achieved is commensurate with
%their corresponding CARs in the vicinity of the posterior mode, rather than
%CARs across the prior as a whole (most clear in the PZ case, compare Table
%\ref{tab:configs} against Figures \ref{fig:pz_acceptance} and
%\ref{fig:pz_acceptance_pmatch}). This is, of course, where a successful chain
%will spend the bulk of its time.

The CAR, introduced in this work, is one means of assessing the impact of
variance in a particle filter's likelihood estimator on the acceptance rate of
a PMMH chain. Computing the CAR at multiple points in parameter space for both
the PZ (Figures \ref{fig:pz_acceptance} and \ref{fig:pz_acceptance_pmatch})
and NPZD (Figure \ref{fig:npzd_acceptcdf}) cases is revealing. For the PZ
model, a clear decline in CAR away from the region of high likelihood is
apparent (Figures \ref{fig:pz_moments}, \ref{fig:pz_acceptance} and
\ref{fig:pz_acceptance_pmatch}), although larger values of the diffusion
parameter $\sigma$ lend improvement. The NPZD case is more complex, with many
parameters having no apparent correlation with CAR over the support of their
prior distribution (Figure \ref{fig:npzd_copulas}). Again, however, there is
some indication that larger values of diffusion parameters (especially $ZDF$)
improve CAR. There is a strong relationship with one parameter, $\mu_{Cl_Z}$,
to which the model is known to be particularly sensitive\footnote{This
  parameter dictates the mean of the stationary distribution of the
  zooplankton clearance rate autoregressive. At low clearance rates,
  phytoplankton will periodically escape zooplankton grazing control and begin
  a rapid bloom, triggering spikes in chlorophyll-a that cannot be reconciled
  with observations. At high clearance rates, phytoplankton is relentlessly
  suppressed by zooplankton predation, keeping chlorophyll-a at much lower
  values than those observed.}.

The dependence of CAR on diffusion parameters is not surprising when the
process model informs the APF proposal distribution: the broader distributions
induced by larger values of diffusion parameters tend to make better
importance proposals, up to a point. For the PZ model, given the uniform prior
over parameters, the maximum \textsl{a posteriori} (MAP) estimate of the
parameters is also the maximum likelihood estimate (MLE) of the parameters,
assuming that the latter falls within the support of the prior. For the NPZD
model, we might assume that the MLE is close to the ground truth.  CAR appears
highest at these MLEs, and declines with distance from them. We conjecture
that this may be a general property, and stress the MLE, not the MAP: the
prior distribution over parameters does not factor into the likelihood
estimator of the particle filter, so the MAP should be relevant only insofar
as it is influenced by the likelihood.

Given the variability of the CAR, there are two potential pitfalls to avoid
when using a PMMH sampler: (i) initialisation in a region where CAR is low,
and (ii) having a particularly informative prior distribution that biases the
posterior into a region where CAR is low. Either case may result in too much
stickiness for the PMMH chain to converge in reasonable time. These should be
considered failure modes of the PMMH sampler in much the same way as strongly
correlated variables can cause slow mixing in the Gibbs sampler, or multiple
modes can cause quasiergodicity in any Markov chain Monte Carlo algorithm. The
CAR is a useful diagnostic for such behaviour.

An alternative approach to PMMH is that of particle
Gibbs~\cite{Andrieu2010}. In place of the Metropolis-Hastings update of
$\boldsymbol{\Theta}$, this involves a Gibbs (or
Metropolis-Hastings-within-Gibbs) update of
$\boldsymbol{\Theta}|\mathbf{u}_{1:T},\mathbf{x}_0,\mathbf{y}_{1:T}$. While
not considered in this work, it is worth noting that the disturbance
state-space model representation may produce better mixing than the
conventional state-space model when sampled with particle Gibbs, because
$\boldsymbol{\Theta}|\mathbf{u}_{1:T},\mathbf{x}_0,\mathbf{y}_{1:T}$ is less
constrained than $\boldsymbol{\Theta}|\mathbf{x}_{0:T},\mathbf{y}_{1:T}$. The
justification is the same as that considered in \citet{Roberts2001}.

\section{Conclusion}\label{sec:conclusion}

In the absence of a closed-form transition density, the disturbance
state-space model seems a generally good approach to enabling cleverer
proposal strategies in the APF. This work establishes some utility in doing
so, particularly for fast-mixing models, by drawing on two specific case
studies in marine biogeochemistry. In these cases the performance of PMMH
chains is shown to improve in some situations by using UKF-based proposals, as
judged by acceptance rate, ESS, convergence rate and CAR. These empirical
results also elucidate some of the behaviours peculiar to the PMMH sampler,
such as the heteroskedasticity of the likelihood estimator, and the implied
need for a good initialisation. The development of robust, generally
applicable and computationally competitive proposal strategies for the APF in
this context remains outstanding work.

\bibliographystyle{abbrvnat}

\appendix

\section{Computing the conditional acceptance rate}\label{app:car}

A simple way to compute the CAR is to first compute the equilibrium
probabilities using (\ref{eqn:p}) and sort them into ascending order. Let
$\mathbf{c}$ be the vector of inclusive prefix-sums over the sorted vector
$\mathbf{p}$:
\begin{equation}
c^i = \sum_{j=1}^i p^j.
\end{equation}
Now proceed as follows:
\begin{equation}
\beta^i = \frac{1}{L}\left[\frac{c^i}{p^i} + L - i\right].
\end{equation}
Substituting into (\ref{eqn:alpha}):
\begin{eqnarray}
CAR(\mathbf{x}_0,\boldsymbol{\theta}) &=& \frac{1}{L}\sum_{i=1}^L p^i \left[\frac{c^i}{p^i} + L - i\right] \\
&=& \frac{1}{L}\left[\sum_{i=1}^L c^i + \sum_{i=1}^L p^i(L - i)\right] \\
&=& \frac{1}{L}\left[\sum_{i=1}^L c^i + \sum_{i=1}^L (c^i - p^i)\right] \\
%&=& \frac{1}{L}\left[\sum_{i=1}^L c^i + \sum_{i=1}^L c^i - \sum_{i=1}^L \tilde{l}^i\right] \\
&=& \frac{1}{L}\left[ 2\sum_{i=1}^L c^i - 1 \right].\label{eqn:car}
\end{eqnarray}

\end{document}